\documentclass[]{JFM-FLM_Au}
\usepackage{wrapfig}
\usepackage{subcaption}
\usepackage{hyperref}
\usepackage{physics}
\usepackage{amsmath}
\newcommand{\upd}{\,\mathrm{d}}
\usepackage[abs]{overpic} 

\lefttitle{M. Bastankhah et al.}
\righttitle{Journal of Fluid Mechanics}

\title{Generalised actuator disk theory: wake development with turbulent entrainment}

\author{Majid Bastankhah\aff{1}, Peter E. Hydon\aff{2},   Carl Shapiro\aff{3}, Dennice F. Gayme\aff{4}, \and Charles Meneveau\aff{4}}

\affiliation{\aff{1}Department of Engineering, Durham University, Durham DH1 3LE, UK
\aff{2} School of Mathematics, Statistics and Actuarial Science, University of Kent, Canterbury CT2 7NF, UK  \aff{3} Pittsburgh, PA, USA \aff{4} Department of Mechanical Engineering, Johns Hopkins University, Baltimore, MD 21218, USA}

\corresau{Majid Bastankhah, \email{majid.bastankhah@durham.ac.uk}}

\hypersetup{
    citecolor=blue,
        }

\begin{document}
\maketitle

\begin{abstract}
Classical actuator disk theory, developed more than a century ago, provides an idealised description of turbine rotor performance. It treats a rotor as an infinitesimally-thin permeable disk and applies the governing flow equations over a streamtube encompassing the disk. A well-known limitation of the theory is its assumption of ideal flow  downstream of the disk, which restricts its applicability to short downwind distances before turbulence and mixing processes governing the wake evolution take hold. The classical theory also leads to unphysical predictions of thrust and power coefficients for highly-loaded rotors. Turbulent axisymmetric wakes,  by contrast, represent an extensively-studied canonical  free shear flow  with much of the  progress and its applications to wind turbines  limited to the far-wake dynamics. In this work, we introduce a generalised actuator disk theory based on a hybrid stream-tube and wake control volume, that seamlessly integrates classical actuator disk analysis with wake turbulence modelling at arbitrary distances from the rotor. The resulting model, while still idealised, can be used to predict variations in velocity, pressure, and cross-sectional flow area as function of position, both upstream and downstream of the rotor disk. Furthermore, by accounting for turbulent entrainment in the wake development, it provides more realistic predictions of thrust and power coefficients for highly-loaded disks.
\end{abstract}


\section{Introduction}
Actuator disk model is one of the most fundamental models in fluid dynamics for analysing turbine rotors and propellers \citep{vanKuik_book2022}. First proposed by Rankine and later refined into its classical form by Froude in 1889 \citep{okulov_betzjoukowsky_2012}, it is now commonly referred to as Froude’s actuator disk theory. The model represents the complex loading of a rotor as a simple pressure jump across an infinitely thin, permeable disk of the same diameter, known as an actuator disk. Despite its simplicity, the theory remains a cornerstone of rotor fluid dynamics because it provides clear and intuitive insights into rotor performance. Froude’s actuator disk theory is typically one of the very first conceptual models introduced to students and researchers interested in the fluid mechanics of turbomachinery.   
It appears in almost every major wind energy textbook \citep[e.g.,][among others]{Burton1995,spera_wind_2009,manwell2010wind,hansen2015aerodynamics,sorensen2016general,branlard2017}, and classical aerodynamics and hydrodynamic textbooks on propellers and rotors \citep[e.g.,][among others]{breslin1994hydrodynamics,leishman2006principles,seddon2011basic,carlton2018marine}.

Despite its historical importance and value as a simple conceptual model, Froude’s theory has important limitations.
First, it does not describe how flow variables evolve with streamwise distance and only relates conditions far upstream and far downstream to those on the actuator disk. Secondly, 
the model does not describe the turbulent wake behaviour 
since the predicted downstream velocity asymptotes to $U_0 (1 - 2a)$, where $U_0$ is the incoming velocity and $a$ is the induction factor, and the wake width becomes constant after an initial expansion.  In reality, following the initial velocity reduction due to pressure recovery, the wake gradually recovers through turbulent mixing, and wake expansion continues downstream.
For these reasons, Froude’s theory is generally regarded as conceptually relevant only in the region immediately downstream of the disk, before turbulence dominates.

The limitations become especially severe for highly-loaded actuator disks (with induction factors $a > 0.5$), where the model even predicts negative wake velocities. Likewise, the well-known relation for the thrust coefficient $C_T = 4a(1-a)$ breaks down in this regime and fails to provide meaningful results \citep{manwell2010wind}. This breakdown is believed to result from strong shear that drives the wake into a fully turbulent state and promotes strong interactions with the ambient flow, effects absent from Froude’s formulation. To mitigate these limitations, empirical corrections are often introduced to adjust the $C_T$–$a$ relationship \citep{Burton1995, buhl2005new}.  For a highly-loaded actuator disk (i.e., a disk with low porosity), where a significant portion of the fluid bypasses the disk rather than passing through it, the actuator disk behaves more like a solid plate. In this case, intense wake turbulence and flow separation generate a strong negative pressure behind the disk, contributing to pressure drag \citep{roshko_wake_1955}. \citet{steiros_drag_2018} accounted for this negative wake pressure by formulating governing equations for a control volume (CV) surrounding the disk that has non-zero pressure at the outlet, unlike Froude's original formulation. The flow inside the CV is assumed to be inviscid and irrotational, and the outlet is placed immediately before the region where turbulent mixing becomes significant. However, because introducing a non-zero outlet pressure adds an extra unknown, additional information is needed to close the system of equations. To address this, they modelled the actuator disk using potential flow theory. As representing the disk as a distribution of potential sources results in an unrealistic velocity discontinuity at the disk plane, they rescaled the wake velocities to enforce mass conservation and validated their predictions against water-channel experiments of porous flat plates. More recently, \citet{liew_unified_2024} adopted a similar CV analysis 
but determined the negative pressure at the CV outlet by solving the two-dimensional pressure Poisson equation proposed by \cite{madsen2023analytical}. They assumed that the CV outlet coincides with the end of the near wake, whose length was estimated using the method of \citet{bastankhah2016experimental}. They also extended this framework to yawed rotors to predict a $C_T$–$a$ relationship that shows good agreement with large-eddy simulation (LES) data for various cases. While these approaches provide improved predictions for highly-loaded disks, they still do not overcome the important limitation of Froude’s actuator disk theory; namely, its inability to incorporate effects of turbulent mixing induced by strong velocity shear that may occur in the immediate downstream vicinity of a highly-loaded disk. 


Turbulent wake development is more realistically represented in far-wake analytical models, which are widely used in wind energy applications to characterise wake interactions in wind farms \citep[see reviews by][and references therein]{stevens2017review,porte2020}. These models are generally based on momentum theory, where pressure effects are neglected under the far-wake assumption, so that the thrust force balances the streamwise flux of momentum deficit  \citep{bastankhah2014new}. Although recent wake models aim to use more realistic velocity profiles in the near wake \citep[e.g.,][among others]{shapiro2019paradigm, blondel2020, schreiber2020brief,ali2024}, their neglect of pressure effects prevents them from accurately capturing the flow immediately behind the disk, where such effects remain significant. Consequently, analytical wake models typically require assumptions about near-wake conditions as inputs. Examples include specifying the onset of the far-wake region, the initial wake width, or the downstream location where the velocity deficit reaches its maximum. Moreover, commonly used far-wake models 
typically
fail to 
provide reliable predictions for induction factors $a > 0.25$ in the far-wake region \citep{bempedelis_analytical_2022}. For instance, the widely used top-hat model of \citet{Frandsen2006} and the Gaussian model of \citet{bastankhah2014new} may predict a \emph{decrease} in the maximum wake velocity deficit as the induction factor $a$ increases beyond 0.25 (i.e., $C_T > 0.75$), which is not physically expected. Although more recent studies \citep[e.g.,][]{bempedelis_analytical_2022} have incorporated pressure effects into the far-wake evolution of highly-loaded disks, they remain limited to the far-wake region, where velocity monotonically increases with 
streamwise distance from the disk. This behaviour is not valid in the region immediately behind the disk.


This paper aims to bridge the gap between two traditionally separate areas of research, actuator disk theory and turbulent wakes. We show that it is possible to overcome the limitations outlined above by developing a new actuator disk theory that includes wake recovery due to turbulent entrainment. While still idealised, the proposed approach provides a more accurate and physically meaningful representation of actuator-disk flows.

\section{Generalised actuator disk theory including wake development}\label{sec:new_actuator_theory}
\begin{figure}
    \centering
    \includegraphics[width=.85\linewidth]{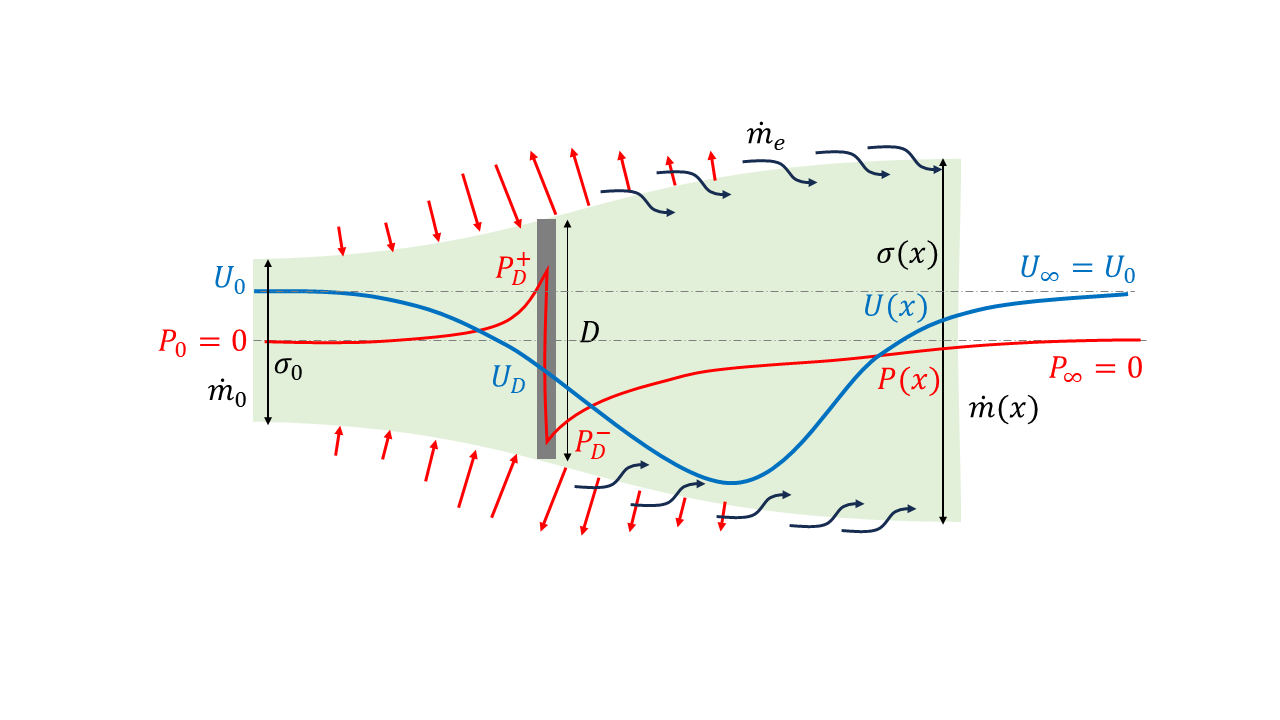}
    \caption{Schematic of the control volume (CV) used in the new proposed actuator disk theory. The upstream part of the CV is a streamtube, while the downstream part follows the wake borders and therefore there is flow entrainment from the lateral area. The outlet is at an arbitrary downstream location with a diameter of $\sigma(x)$, pressure $P=P(x)$ and velocity $U=U(x)$. The disk is located at $x=0$. }
    \label{fig:schematic}
\end{figure}

We consider the ``hybrid'' CV around a disk as shown in Figure~\ref{fig:schematic}. Upwind of the disk, as in Froude’s theory, we assume there is no mass exchange between the CV and the surroundings, so the CV is the streamtube encompassing the rotor disk. This assumption implies  that turbulent mixing does not play a significant role in shaping the flow distribution upstream of the disk. This notion is supported by previous studies showing that inviscid solutions often provide satisfactory predictions in this region \citep[e.g.,][]{medici2011upstream,Bastankhah2017POF,segalini_analytical_2021}.  Downstream of the rotor disk, however, the CV is not a streamtube but is assumed to coincide with the outer boundary of the wake, where turbulence drives entrainment of the ambient flow through the lateral surface of the CV, until the wake is fully recovered.
For this CV, at each streamwise position $x$, the velocity $U(x)$ and pressure $P(x)$ are assumed to be uniform, with the diameter of the CV cross-section denoted by $\sigma(x)$. The disk is located at $x=0$, and the diameter of the CV at $x=0$ is equal to the disk diameter $D$.
Subscripts $0$ and $\infty$ denote the asymptotic far-upstream ($x\!\to\!-\infty$) and far-downstream ($x\!\to\!\infty$) values, respectively.

In Froude’s original formulation, the CV extends from far upstream to far downstream. Here, to capture the streamwise variations of the flow quantities, we retain the inlet at far upstream but place the CV outlet at an arbitrary streamwise location $x$, where $x$ can take on positive or negative values.
The goal of our analysis is to determine $P(x)$, $U(x)$, $\sigma(x)$, and thrust coefficient $C_T$ for a given induction factor $a$ (or for a known $C_T$, to determine $a$).

\subsection{Mass conservation}
The conservation of mass for this CV is written as
\begin{equation}\label{eq:continuity_new}
    \dot{m}(x)=\dot{m}_0+\dot{m}_e(x),
\end{equation}
where $\dot{m}_0=\frac{\pi}{4} \rho \sigma^2_0 U_0$ and $\dot{m}(x)=\frac{\pi}{4} \rho \sigma^2(x) U(x)$ are the mass flow rates at the inlet and outlet, respectively, and $\dot{m}_e(x)$ represents the mass flow rate of the entrained flow across the lateral surface. Also,  $\rho$ is the air density.  If we define the entrainment velocity as $U_e(x)$,   the axial rate of change of $\dot{m}_e(x)$ can be written as the product of perimeter times entrainment velocity and  is given by
\begin{equation}\label{eq:entrainment_mass}
    \frac{\mathrm{d}\dot{m}_e}{\mathrm{d}x}=\rho\pi \,\sigma(x) \, U_e(x) \, \mathrm{H}(x),
\end{equation}
 where $\mathrm{H}(x)$ is the Heaviside function, which ensures that the entrainment driven by turbulence only takes place downwind of the actuator disk. 
  Taking the derivative of equation~\eqref{eq:continuity_new} with respect to $x$ and using equation~\eqref{eq:entrainment_mass}, we obtain
\begin{equation}\label{eq:sigma2u_new}
    \frac{\mathrm{d}\left(\sigma^2 U\right)}{\mathrm{d}x}=4\sigma U_e\mathrm{H}(x),
\end{equation}
which can be expanded and rearranged as
\begin{equation}\label{eq:dsigmadx}
    \frac{\mathrm{d}\sigma}{\mathrm{d}x} = \frac{1}{2U} \left( - \sigma\frac{\mathrm{d}U}{\mathrm{d}x} + 4U_e\mathrm{H}(x)\right).
\end{equation}
\subsection{Choice of entrainment velocity $U_e$}
\subsubsection{Wake-shear driven entrainment} 
 In the absence of ambient turbulence, turbulent entrainment is primarily driven by wake shear. In studies applying entrainment theory to wake flows \citep{morton_momentum-mass_1961,luzzatto2018,luzzatto-fegiz_one-parameter_2018, bempedelis2023,bempedelis_analytical_2022}, the entrainment rate is typically modelled as a function of the characteristic velocity difference, so the entrainment velocity is modelled as
    \begin{equation}\label{eq:entrainment_wake_shear}
        \frac{U_e^w(x)}{U_0} = E_1 \frac{\Delta U(x)}{U_0},
    \end{equation}
    where $\Delta U(x) = U_0 - U(x)$ represents the velocity deficit, the superscript $w$ denotes wake-shear driven entrainment, and the parameter $E_1$ is the shear-driven entrainment coefficient. Here, we assume $E_1 = 0.1$, consistent with classical entrainment studies \citep{morton_momentum-mass_1961}.  

\subsubsection{Background-turbulence driven entrainment}\label{sec:entrainment_background}
In addition to wake shear, background turbulence also plays a significant role in the entrainment process and wake evolution. This is particularly relevant for cases where the actuator disk (e.g., a wind turbine) operates in a deep atmospheric boundary layer that contains large-scale energetic eddies. These eddies typically have a cross-stream length scale comparable to the height from the ground, and often a streamwise length considerably larger than the wake width $\sigma$, scaling closer to the overall boundary layer thickness. We denote their overall size by an integral length scale $\Lambda$ in this work.  
Under these conditions, the wake is transported and meandered by the incoming large-scale turbulent eddies \citep{larsen2008}. Therefore, the entrainment velocity in this regime is determined primarily by the inflow characteristics. We assume that the entrainment velocity driven by background turbulence is proportional to the root-mean-square of the incoming streamwise velocity fluctuations. When normalised by the incoming velocity $U_0$, this implies that the normalised entrainment velocity is proportional to the streamwise turbulence intensity $I$. It is worth noting that the dependency of the entrainment velocity on the integral length scale $\Lambda$ is neglected here for simplicity, which may not be completely true \citep{hodgson_effects_2023, vahidi_influence_2024,li_impacts_2024}. 
We further note that the turbulence intensity can be defined based on either the total turbulent kinetic energy or its streamwise component alone. While the former may be more physically grounded, the latter is easier to measure and is the standard definition used in most current wind energy research \citep{porte2020}. There is a subtle point regarding the onset of entrainment due to background turbulence that also needs to be discussed. Previous studies \citep[e.g.][]{lignarolo2015tip} have demonstrated that the strong vortex layer surrounding the near wake inhibits interaction with the ambient flow. To capture this shielding effect, we link the entrainment contribution from background turbulence to the degree of pressure recovery downstream of the wake. This implies that the entrainment due to background turbulence is effectively suppressed at $x=0$ and asymptotically approaches its maximum value as the wake pressure recovers. Inspired by the pressure variation downstream of the actuator disk derived later in section~\ref{sec:pressure_variation} (equation~\eqref{eq:PPE_solution_semiinf_f=g=0}), we propose the following relation:
\begin{equation}
\frac{U_e^{b}}{U_0} =  E_2 I \frac{x}{\sqrt{x^2+R^2}},
\label{eq:entrainment-atmospheric}
\end{equation}
where $E_2$ is the background-driven entrainment coefficient, superscript $b$ stands for background turbulence, and $R=D/2$ is the rotor radius.  It will be shown later in section \ref{sec:far_wake_ambient} that $E_2$ is expected to be in the range of $[0.25-1]$. In this study, we use the value $E_2=0.6$ as it provides reasonable agreement with experimental and LES data presented later in section~\ref{sec:validation}.

\subsubsection{Total entrainment driven by wake shear and background turbulence} 
We  propose that the total entrainment velocity is proportional to the generalised mean of the normalized velocity deficit and the incoming turbulence intensity:
\begin{equation}
\frac{U_e(x)}{U_0} = \left[ \left( \frac{U_e^w}{U_0}\right)^n + \left( \frac{U_e^{b}}{U_0} \right)^n \right]^{1/n}
\label{eq:entrainment_velocity}
\end{equation}
where $n$ is a positive integer. Here, we assume $n=4$, noting that larger values do not lead to a considerable difference. The advantage of using equation~\eqref{eq:entrainment_velocity} is that the entrainment velocity closely approaches the larger of the two contributing factors. Consequently, under typical conditions, the near-wake entrainment velocity is dominated by the wake shear, which is generally larger in that region, while in the far wake the entrainment velocity is dominated by  background turbulence.  

\subsection{Momentum balance}\label{sec:conservation of momentum}
\begin{figure}
    \centering
    \begin{subfigure}[b]{0.18\textwidth}
        \centering
        \includegraphics[width=\textwidth]{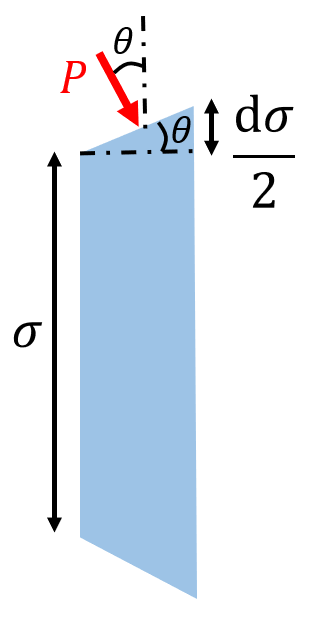}
        \caption{}
        \label{fig:side_pressure}
    \end{subfigure}
    \hspace{1cm} 
    \begin{subfigure}[b]{0.5\textwidth}
        \centering
        \includegraphics[width=\textwidth]{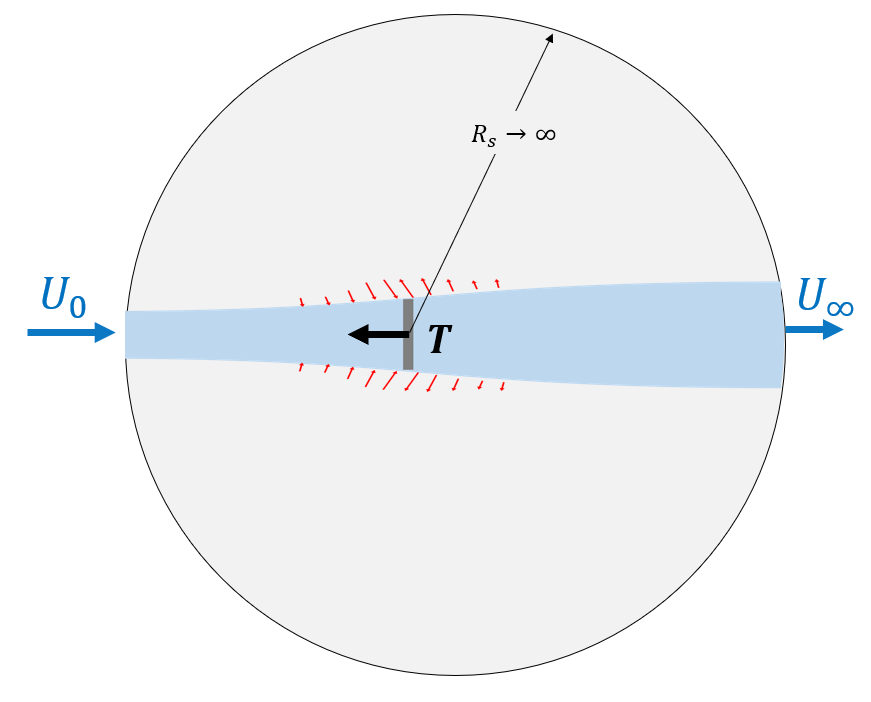}
        \caption{}
        \label{fig:spherical_CV}
    \end{subfigure}
    \caption{(a) Pressure force exerted on an infinitesimal lateral area. (b) Schematic of a spherical CV with an infinitely large radius surrounding the disk. For this CV, unlike the one shown in figure~\ref{fig:schematic} that is used in our present model formulation, lateral pressure forces are internal forces and do not appear in the momentum equation.}
    \label{fig:combined_figure}
\end{figure}

The momentum equation for the CV shown in figure \ref{fig:schematic} is given by
\begin{equation}\label{eq:momentum}
 \rho\frac{\pi}{4}\sigma^{2}(x)\, U(x) \, [U_{0}-U(x)] =\rho\frac{\pi}{8} C_{T}U_{0}^{2} D^{2} \mathrm{H}(x)+\frac{\pi}{4}\sigma^{2}(x) P(x) -F_{P_s}(x), 
\end{equation}
where the momentum deficit flux term on the left-hand-side is written using the mass conservation (equation \eqref{eq:continuity_new}), i.e. 
$\dot{m}_e=\frac{\pi}{4} \rho (\sigma^2 U- \sigma^2_0 U_0)$ and its associated momentum flux $\dot{m}_e U_0$. 
In equation, \eqref{eq:momentum} the term $F_{P_s}(x)$ is the axial component of the force exerted by the pressure on the lateral surface of the streamtube, given by
\begin{equation}\label{eq:F_p_s}
    F_{P_s}(x)= \frac{\pi}{4}\int_{-\infty}^{x}P(x') \, \frac{\upd \sigma^2(x')}{\upd x'} \upd x',
\end{equation}
where 
$x'$ is a  dummy variable. Equation~\eqref{eq:F_p_s} is obtained by integrating the pressure force exerted on an infinitesimal lateral area, shown in figure \ref{fig:side_pressure}, from far upstream to $x$.  In Froude’s actuator theory, the lateral pressure term is neglected because the CV's outlet is assumed to be sufficiently far downstream, such that the lateral pressure force contribution downstream cancels that upstream, resulting in a net effect of zero. This zero net contribution in the far wake can be understood by considering a spherical control volume of very large radius, $R_s\!\to\! \infty$, surrounding the disk, as shown in figure~\ref{fig:spherical_CV}. For this CV, the lateral pressure forces are all internal \citep{vanKuik_book2022}, and the pressure at the control surface is atmospheric everywhere. Consequently, equation~\eqref{eq:momentum} reduces to a simple balance between the thrust force and the net momentum flux across the control surface. However, it is  important to note that for a finite $x$ (either positive or negative) considered here, the net contribution of the lateral pressure term is non-zero and must be retained in the momentum equation.

Taking the derivative of equation \eqref{eq:momentum} with respect to $x$ and rearranging terms leads to
\begin{equation}\label{Eq:mom_new}
   U\frac{\mathrm{d}U}{\mathrm{d}x}+ \frac{1}{\rho}\frac{\mathrm{d}P}{\mathrm{d}x}=\left(\frac{2U}{\sigma}\frac{\mathrm{d}\sigma}{\mathrm{d}x}+\frac{\mathrm{d}U}{\mathrm{d}x}\right)(U_0-U)-\frac{1}{2} C_{T}U_{0}^{2} \mathrm{\delta}(x),
\end{equation}
where $\delta(x)$ is the Dirac delta function. 
Using equation \eqref{eq:dsigmadx} to express $\upd\sigma/\upd x$ allows equation \eqref{Eq:mom_new} to be written as
\begin{equation}\label{eq:mom_new2}
       U(x)\frac{\mathrm{d}U}{\mathrm{d}x}+ \frac{1}{\rho}\frac{\mathrm{d}P}{\mathrm{d}x}=-\frac{1}{2} C_{T}U_{0}^{2} \mathrm{\delta}(x) + \frac{4U_e(x)[U_0-U(x)]\, \mathrm{H}(x)}{\sigma(x)}.
\end{equation}
It is worth noting that \eqref{eq:mom_new2} bears a strong resemblance to the differential Reynolds-Averaged Navier-Stokes (RANS) equation in the $x$-direction, provided the Reynolds shear stress term is simplified using scaling arguments (i.e., replacing radial gradients with variations over wake radius $\sigma/2$). This comparison also suggests that $U_e\sigma$ essentially acts as an effective turbulent viscosity. Integrating \eqref{eq:mom_new2} from $-\infty$ to $x$  results in the following Bernoulli-type equation
\begin{equation}\label{eq:Bernoulli_modified2}
    \frac{1}{\rho}\,P(x)+\frac{1}{2}U^2(x)=\frac{1}{2}U_0^2\left(1- C_T \mathrm{H}(x)  \right)+  4 \int_0^{\mathrm{max}(x,0)} \frac{U_e(x')\,[U_0-U(x')]}{\sigma(x')} \, \mathrm{d}x', 
\end{equation}
which represents the variation of the mechanical energy with $x$. It can also be regarded as a generalised form of the Bernoulli equation, with two additional effects incorporated: the extraction of energy by the disk (i.e., $-C_T\mathrm{H}(x)$ term on the right-hand side) and the entrainment of energy by turbulence (the last term on the right-hand side). When $C_T=U_e = 0$, this equation reduces to a Bernoulli equation. It is also important to note that the derivation of this generalised form of Bernoulli equation from mass and momentum equations is possible when the momentum equation is written in its complete form including the lateral pressure term $F_{P_s}$.



Finally, equation~\eqref{eq:mom_new2} can be rearranged as 
\begin{equation}\label{eq:dudx}
\frac{\mathrm{d}U}{\mathrm{d}x} = \frac{1}{U} \left( -\frac{1}{\rho} \frac{\mathrm{d}P}{\mathrm{d}x} - \frac{1}{2} C_{T} U_0^2\, \delta(x) + \frac{4U_e(U_0 - U)\, \mathrm{H}(x)}{\sigma} \right).
\end{equation}
We then have a system of two ordinary differential equations comprising equation~\eqref{eq:dudx} for $\mathrm{d}U/\mathrm{d}x$ and equation~\eqref{eq:dsigmadx} for $\mathrm{d}\sigma/\mathrm{d}x$. However, there are three dependent variables, $U(x)$, $\sigma(x)$, and $P(x)$, with $x$ as the independent variable. The required third equation (for $\mathrm{d}P/\mathrm{d}x$) is obtained in the next section.

\subsection{Pressure variations for lightly-loaded actuator disks}\label{sec:pressure_variation}
Here, we aim to solve a simplified form of the pressure Poisson equation to determine $P(x)$. Previous studies have solved the linearised pressure Poisson equation over the entire domain ($-\infty\! <\! x\! <\! \infty$) that includes actuator-disk forcing using 3D axisymmetric \citep{koning_influence_1935, segalini_analytical_2021} and 2D Cartesian \citep{madsen_analytical_2023,liew_unified_2024} formulations, although 2D approaches are inherently less realistic for 3D rotors. Since linearised solutions neglect non-linear terms in the pressure Poisson equation, they yield pressure jumps across the rotor disk that are strictly valid only for small values of induction factor. To overcome this limitation, \cite{madsen_analytical_2023} used a scaled $C_T$ to match the pressure jump obtained from the Poisson equation to the one obtained from Bernoulli equation, while \cite{liew_unified_2024} solved the pressure Poisson equation numerically to retain the effects of non-linear terms. In this study, we take a different approach. We solve the linearised pressure Poisson equation separately for the upwind and downwind half-spaces with no actuator-disk forcing, which effectively simplifies it to Laplace's equation. By enforcing the correct pressure jump, obtained from the Bernoulli equation \eqref{eq:Bernoulli_modified2} as a boundary condition, we reduce the error caused by omitting the non-linear terms, 
at least in the region immediately upstream and downstream of the actuator disk.

\begin{figure}
    \centering
    \includegraphics[width=0.6\linewidth]{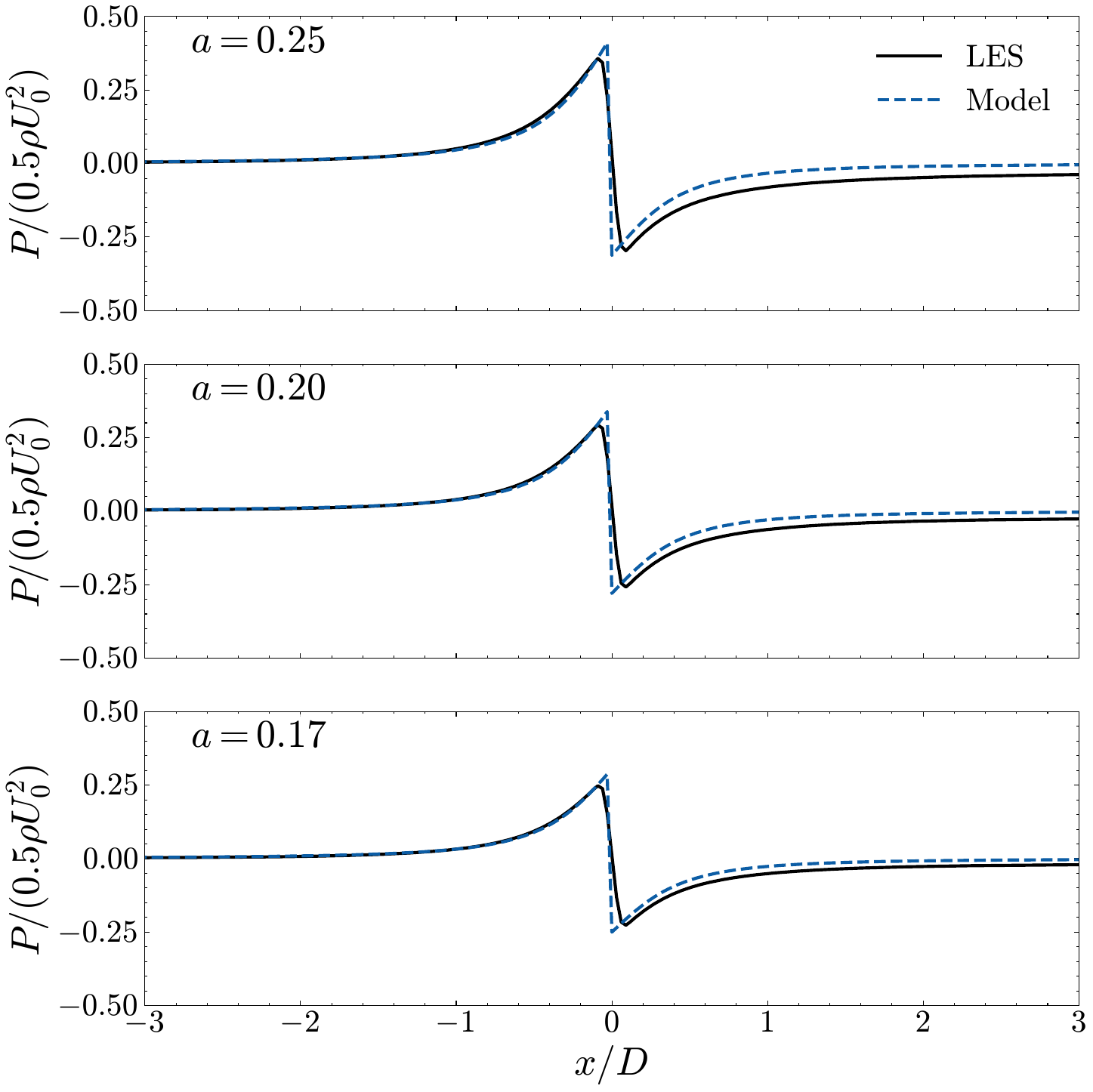}
    \caption{Comparison of the predictions of equation \eqref{eq:PPE_solution_semiinf_f=g=0} with  LES data under laminar inflow conditions.}
    \label{fig:pressure_laminar}
\end{figure}

As discussed above, for a lightly-loaded actuator disk (i.e. small values of the induction factor, $a$), the axisymmetric pressure Poisson equation can be simplified to the Laplace equation $\nabla^2P=0$ as a leading-order approximation (see Appendix \ref{sec:appendix} for details). In the half-space upstream of the rotor disk the Dirichlet boundary condition on the disk $P(x=0^-)=P^+_D$ is used, while on the downstream half-space one uses $P(x=0^+)=P^-_D$, where $P_D^{+}$ and $P_D^{-}$ are the values of pressure immediately upstream and downstream of the actuator disk, respectively. Along the centreline,
the solution to the Laplace equation is given by
\begin{equation}\label{eq:PPE_solution_semiinf_f=g=0}
   P(x)=P_D^{+}\left(1+\frac{x}{\sqrt{x^2+R^2}}\right)\mathrm{H}(-x)+P_D^{-}\left(1-\frac{x}{\sqrt{x^2+R^2}}\right)\mathrm{H}(x).
\end{equation}
The interested reader is referred to Appendix \ref{sec:appendix} for the complete mathematical solution of the pressure Poisson equation (including non-linear terms), and underlying assumptions and simplifications leading to equation \eqref{eq:PPE_solution_semiinf_f=g=0}. From equation \eqref{eq:Bernoulli_modified2} with $U_D=U_0(1-a)$, $P_D^{+}$ and $P_D^{-}$ are respectively given by  
\begin{equation}\label{eq:P_D}
P_D^+ = \frac{1}{2}\rho U_0^2a(2-a), \quad P_D^- = \frac{1}{2}\rho U_0^2\left(a(2-a) - C_T\right).    \end{equation}
We compare $P(x)$ from equation \eqref{eq:PPE_solution_semiinf_f=g=0} with those from numerical simulations. The interested reader is referred to  \cite{shapiro2018modelling} and \cite{bastankhah_vortex_2022} for more information on the numerical setup, which consists of a LES  with laminar uniform inflow towards an actuator disk, and negligible effects of turbulence in the near-rotor region. Results are shown in figure \ref{fig:pressure_laminar} for three different values of $a$. The model (dashed lines) requires specification of $C_T$ for the downstream portion of the flow. As will be shown later in \S \ref{sec:determine ct}, for the case $U_e=0$, the classical Froude relation $C_T=4a(1-a)$ holds. Since for $x/D<3$ the effects of turbulence were negligible in the simulations, we use $C_T=4a(1-a)$ to specify $C_T$ in this comparison with idealised simulation results. Results shown in  figure \ref{fig:pressure_laminar}  confirm that equation~\eqref{eq:PPE_solution_semiinf_f=g=0} yields satisfactory predictions of the pressure distribution, with accuracy improving as \( a \) decreases.  
It is worthwhile mentioning that, according to equation \eqref{eq:P_D}, the pressure drop across the actuator disk is not necessarily symmetric. In fact, for typical values of $C_T$, the magnitude of the immediate upstream pressure increase, $|P_D^+|$, is greater than the immediate downstream pressure drop, $|P_D^-|$. This subtle fact is also important for determining the pressure variations 
(see Appendix \ref{sec:appendix} for details).

Taking the derivative of equation \eqref{eq:PPE_solution_semiinf_f=g=0} and inserting values of $P_D^{+}$ and $P_D^{-}$ from equation \eqref{eq:P_D} gives
\begin{equation}\label{eq:dp/dx}
   \frac{\upd P}{\upd x} = \frac{1}{2} \rho U_0^2 \left( \frac{R^2}{(x^2 + R^2)^{3/2}} \left[ a(a - 2) \operatorname{sgn}(x) + C_T\,\mathrm{H}(x) \right] - C_T \, \delta(x) \right),
\end{equation}
where \( \operatorname{sgn}(x) \) is the sign function, defined as
$\operatorname{sgn}(x) = -1$ if $x<0$, and $\operatorname{sgn}(x) = 1$
if $x>0$. We now have the three relations needed to obtain the streamwise evolution of velocity $U(x)$, pressure $P(x)$, and CV diameter $\sigma(x)$ for given values of the induction factor $a$ and thrust coefficient $C_T$.


\subsection{Solving for $U(x)$, $\sigma(x)$, and $P(x)$}
\label{sec:determine U and sigma}
Here, we determine $\sigma(x)$, $U(x)$, and $P(x)$ for given $C_T=C_T(a)$. Inserting $\upd P/\upd x$ from \eqref{eq:dp/dx} in equation \eqref{eq:dudx} and using \eqref{eq:dsigmadx}, we obtain a system of equations that can be solved to find $U(x)$ and $\sigma(x)$
\begin{subequations}\label{eq:sigma_u_final}
\begin{align}
\frac{\mathrm{d}\sigma}{\mathrm{d}x} &= \frac{1}{2U} \left( - \sigma\frac{\mathrm{d}U}{\mathrm{d}x} + 4U_e\mathrm{H}(x) \right), \label{eq:sigma_u_final_a} \\[6pt]
\frac{\mathrm{d}U}{\mathrm{d}x} &= \frac{U_0^2}{2U} \left( 
\frac{R^2}{(x^2 + R^2)^{3/2}} \left( a(2-a)\, \operatorname{sgn}(x) - C_T \, \mathrm{H}(x) \right)
+ \frac{8}{\sigma}\frac{U_e}{U_0}\frac{U_0 - U}{U_0}\, \mathrm{H}(x)
\right). \label{eq:sigma_u_final_b}
\end{align}
\end{subequations}
For the upwind region ($x<0$), equation \eqref{eq:sigma_u_final} is readily solved, yielding
\begin{subequations}\label{eq:upwind_solution}
\begin{align}
\frac{U(x<0)}{U_0} &= \sqrt{a(a-2)\left(\frac{x}{\sqrt{x^2+R^2}}+1\right)+1}, \label{eq:upwind_solution_a} \\[6pt]
\frac{\sigma(x<0)}{D} &= \sqrt{\frac{U_0(1-a)}{U(x)}} 
= \sqrt{1-a} \,\left[ a(a-2)\left(\frac{x}{\sqrt{x^2+R^2}}+1\right)+1 \right]^{-1/4}. \label{eq:upwind_solution_b}
\end{align}
\end{subequations}

For the downwind region ($x>0$), no analytical solution was found due to the presence of $U_e(x)$. Instead, a basic forward marching numerical scheme (or more efficient Runge–Kutta methods) can be used, starting at $x=0$. The initial values   $U(0)=U_0(1-a)$ and $\sigma(0)=D$ can be used to solve equation \eqref{eq:sigma_u_final} numerically and compute $U(x)$ and $\sigma(x)$. Pressure $P(x)$ is directly obtained from equation \eqref{eq:PPE_solution_semiinf_f=g=0}. 

\subsection{Determining the thrust coefficient $C_T$}\label{sec:determine ct}
So far, we have not determined a general relationship between $a$ and $C_T$. 
Next, we  apply the proposed generalised actuator disk model with turbulent entrainment to develop a relationship for $C_T$ and establish a framework in which all three flow quantities ($U,\sigma$, and $P$) and $C_T$ are determined for a given value of $a$. The obtained relations can also be inverted to obtain $a$ for any prescribed $C_T$, since it is the latter quantity that is generally known a-priori for a given rotor. 

Section \ref{sec:conservation of momentum} discussed that the net contribution of the lateral pressure term in the momentum equation must asymptote to zero as $x\!\to\!\infty$. By enforcing this condition, a new relation for $C_T$ can be obtained. 
The total force from the side pressure contribution is given by $F_{P_s}(x\to \infty)$ from equation \eqref{eq:F_p_s} and implies: 
\begin{equation}\label{eq:F_p_s2}
     \frac{\pi}{4}\int_{-\infty}^{0} P(x) \, \frac{\upd \sigma^2(x)}{\upd x} \upd x +  \frac{\pi}{4}\int_{0}^{\infty}P(x) \, \frac{\upd \sigma^2(x)}{\upd x} \upd x  = 0.
\end{equation}
Denoting the first and second integrals as $F_{P_s}^+$ and $F_{P_s}^-$, respectively, we can evaluate the upwind  lateral pressure force as
\begin{equation}\label{eq:f_ps+}
    F_{P_s}^+=\frac{\pi}{4}\int_{-\infty}^{0} P(x) \, \frac{\upd \sigma^2(x)}{\upd x} \upd x  = \frac{\pi}{8}\rho D^2U_0^2a^2,
\end{equation}
where we have used equations \eqref{eq:PPE_solution_semiinf_f=g=0} and \eqref{eq:P_D} for $P(x)$ and \eqref{eq:upwind_solution_b} for $\sigma(x)$.
From equations \eqref{eq:F_p_s2} ($F_{P_s}^- = - F_{P_s}^+$) and \eqref{eq:f_ps+}, we conclude that
\begin{equation}\label{eq:f_ps+=f_ps-}
   \int_0^{\infty}P(x)  \frac{\upd \sigma^2(x)}{\upd x}\upd x= -\frac{1}{2}\rho D^2U_0^2a^2,
\end{equation}
which can be expanded using equations \eqref{eq:PPE_solution_semiinf_f=g=0} and \eqref{eq:P_D} to
\begin{equation}
    \int_0^{\infty}\left(a(2-a) - C_T\right)\left(1-\frac{x}{\sqrt{x^2+R^2}}\right)\frac{\upd \sigma^2(x)}{\upd x}\upd x=-D^2a^2.
\end{equation}
Solving for $C_T$ gives
\begin{equation}\label{eq:CT_new}
    C_T=2a+\left(\frac{1}{Y}-1\right)a^2,
\end{equation}
where
\begin{equation}\label{eq:Y}
    Y=\frac{1}{D^2}\int_0^{\infty}\left(1-\frac{x}{\sqrt{x^2+R^2}}\right)\frac{\upd \sigma^2(x)}{\upd x}\upd x.
\end{equation}
To solve equation \eqref{eq:CT_new} with $Y$ given by equation \eqref{eq:Y}, $\sigma(x)$ is required. However, from equation \eqref{eq:sigma_u_final}, determining $\sigma(x)$ itself requires knowledge of $C_T$. Consequently, equations \eqref{eq:sigma_u_final} and \eqref{eq:CT_new} must be solved iteratively. We begin by guessing a value for $C_T$ (e.g., using $C_T = 4a(1-a)$ or other models such as \citet{buhl2005new,steiros_drag_2018,liew_unified_2024}) and use it in equation \eqref{eq:sigma_u_final} to compute $\sigma(x)$. The resulting $\sigma(x)$ is then substituted into equations \eqref{eq:Y} and \eqref{eq:CT_new} to obtain an updated $C_T$. This updated value is fed back into equation \eqref{eq:sigma_u_final}, and the process is repeated until convergence is achieved. 

If instead of $a$, we are provided with $C_T$,  we begin by guessing a value of $a$ and use it in equation \eqref{eq:sigma_u_final} to compute $\sigma(x)$. The resulting $\sigma(x)$ is then substituted into equation \eqref{eq:Y} to find the value of $Y$, which is inserted in equation \eqref{eq:CT_new} to find a new value for $a$, and the process is repeated until convergence is achieved. Implementations of these iterative procedures can be found in the computational notebooks associated with figures \ref{fig:different_a} or \ref{fig:ct_cp} (see the figure captions for details).  

It is worth noting that the integral $Y$ in equation~\eqref{eq:Y} converges only if the wake width scales as $\sigma \propto x^{\alpha}$ with $\alpha < 1$. 
The choice of a constant background entrainment velocity $U_e^b$ results in a linear asymptotic wake expansion, which technically causes the integral to slowly diverge. Physically, however, the wake cannot grow linearly indefinitely. For wind turbines, its expansion is eventually constrained by the ground. More importantly, the underlying assumption that $\Lambda \gtrsim \sigma$ becomes invalid at very large streamwise distances. 
In light of these considerations, we truncate the integration in equation~\eqref{eq:Y} at $x=3D$, where according to equation~\eqref{eq:PPE_solution_semiinf_f=g=0}, approximately 99\% of the pressure recovery is already achieved by this distance.

\subsection{Reduction to Classical Froude Theory for $U_e=0$}
Solving equation \eqref{eq:sigma_u_final} for $U_e=0$ at $x>0$, we obtain 
\begin{equation}\label{eq:downind_solution_ue=0}
   \frac{U(x>0)}{U_0}= \sqrt{(1-a)^2+\big[(2a-a^2)-C_T\big]\frac{x}{\sqrt{x^2+R^2}}},
\end{equation}
and $\sigma(x)=D\sqrt{U_0(1-a)/U(x)}$. For the upwind region ($x<0$), the solution is the same as equation \eqref{eq:upwind_solution}. From equation \eqref{eq:downind_solution_ue=0}, the asymptotic velocity   at $x\to\infty$ is $U_{\infty}=U_0\sqrt{1-C_T}$, or $U_{\infty}=U_0(1-2a)$ if $C_T=4a(1-a)$. The well-known relation of $C_T=4a(1-a)$ can be also derived from our new generalised theory for $U_e=0$. Inserting equation~\eqref{eq:downind_solution_ue=0} into equation~\eqref{eq:Y} yields 
\( Y = -1 + \tfrac{2(1-a)}{\,a(2-a)-C_T\,}\left(\sqrt{1-C_T} - (1-a)\right) \), 
which can then be substituted into equation~\eqref{eq:CT_new}, whose only non-trivial solution becomes 
\( C_T = 4a(1-a) \).

\section{Results and discussions}

\begin{figure}
    \centering
    \includegraphics[width=.9\linewidth]{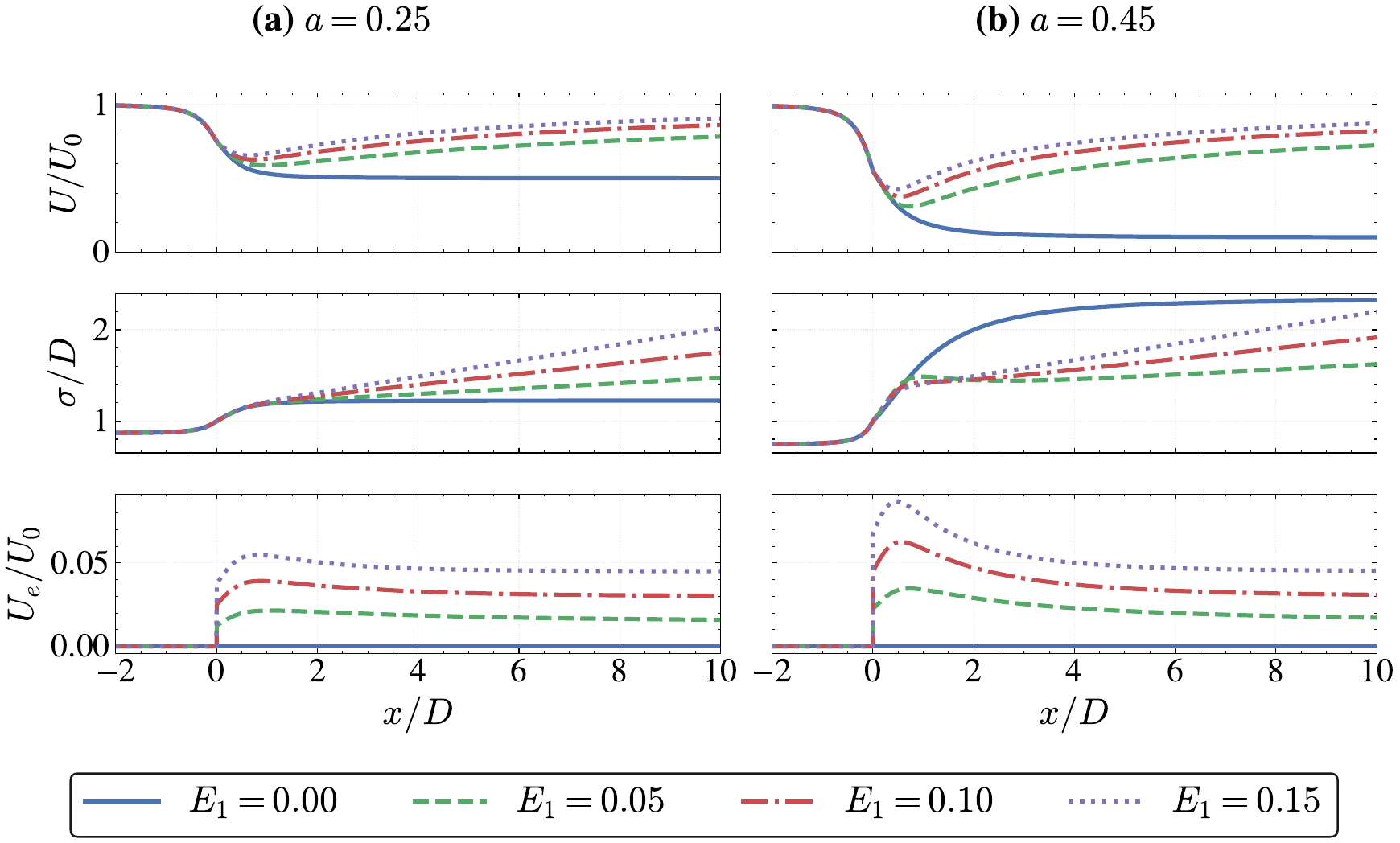}
    \caption{Variations of flow properties for different entrainment coefficients $E_1$ ($E_2 = 0.6$, $I = 0.05$) at two axial induction factors: (a) $a = 0.25$ and (b) $a = 0.45$.}
    \label{fig:different_E}
\end{figure}

\subsection{Model predictions and trends}\label{sec:3-1}

The influence of entrainment velocity on model predictions is first illustrated in figure~\ref{fig:different_E}, which presents results for various values of the entrainment coefficients $E_1$, with $E_2 = 3E_1$ and $I = 5\%=0.05$. Results are presented for two different values of induction factor $a$ of $0.25$ and $0.45$. When $E_1 =E_2= 0$, there is no turbulent entrainment, and the far-wake velocity approaches $U_0(1 - 2a)$, as shown in figure \ref{fig:different_E}. As expected, with increasing the level of turbulent entrainment, the wake velocity recovers more rapidly for both cases shown. However, the effect of entrainment on $\sigma$ in the near wake differs between the two induction factors. At $x < 2.5D$, turbulent entrainment slightly increases $\sigma$ for the lower induction factor ($a=0.25$), whereas for the higher induction factor ($a = 0.45$) turbulent entrainment significantly reduces the near-wake width compared to the case without turbulence. This behaviour can be interpreted using equation~\eqref{eq:sigma_u_final_a}. According to this relation, $\sigma$ in the near-wake region may increase due to two mechanisms: (i) turbulent entrainment (second term on the right-hand side, $4U_e\mathrm{H}(x)$), and (ii) flow deceleration occurring behind the disk (first term on the right-hand side, $-\sigma \,\upd U/\upd x$). However, as shown in figure~\ref{fig:different_E}, flow deceleration behind the disk is less in the presence of turbulent entrainment. Consequently, it has opposing effects on $\sigma$, and the net outcome, whether $\sigma$ increases or decreases, depends on which one dominates.

\begin{figure}
    \centering
    \includegraphics[width=0.8\linewidth]{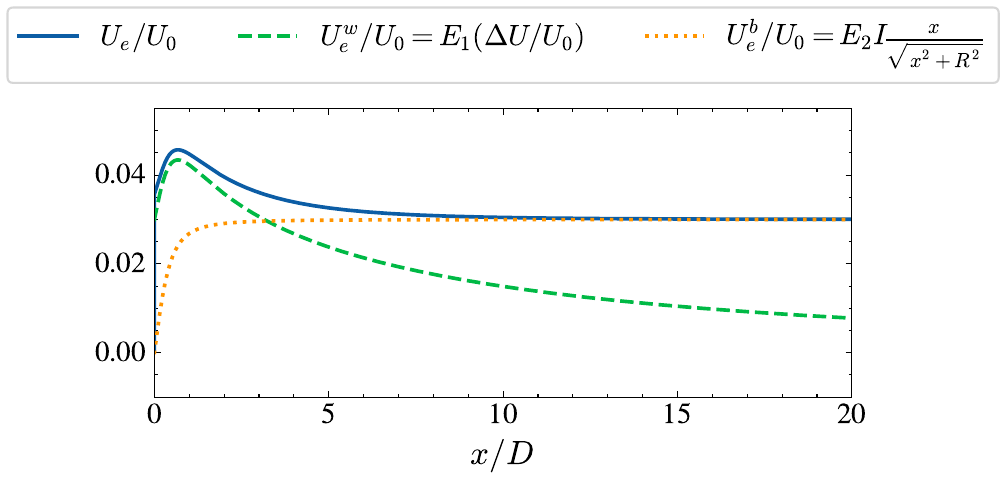}
    \caption{Variations of entrainment velocity with downwind distance, where $U_e$ is the total entrainment velocity (equation \eqref{eq:entrainment_velocity}), $U_e^w$ is the wake-shear driven entrainment velocity (equation \eqref{eq:entrainment_wake_shear}), and $U_e^{b}$ is the background-turbulence driven entrainment velocity (equation \eqref{eq:entrainment-atmospheric}).}
    \label{fig:entrainemnt_velocity}
\end{figure}

Figure~\ref{fig:entrainemnt_velocity} shows how the entrainment velocity varies with the streamwise distance downwind of the actuator disk for a typical case with $a = 0.3$ and $I = 5\%$, using values of $E_1 = 0.1$ and $E_2 = 0.6$. According to equation~\eqref{eq:entrainment_velocity}, the total entrainment velocity $U_e$ consists of two components: (i) the wake-shear driven entrainment velocity $U_e^w$ (equation \eqref{eq:entrainment_wake_shear}), and (ii) the background-turbulence-driven entrainment velocity $U_e^b$ (equation \eqref{eq:entrainment-atmospheric}). The figure shows that turbulent entrainment is predominantly driven by wake shear in the region immediately downstream of the actuator disk. At intermediate streamwise distances (e.g., $2D\!< \! x \! < \! 7D$), the entrainment process is governed by a combination of wake shear and background turbulence. However, in the far-wake region ($x > 7D$), background turbulence becomes the dominant mechanism driving wake recovery. The precise location of the transition between these regimes depends on the thrust coefficient $C_T$ and ambient turbulence intensity $I$ \citep{bastankhah2016experimental}, as well as the 
values chosen for the coefficients $E_1$ and $E_2$.

\begin{figure}
    \centering
    \includegraphics[width=1\linewidth]{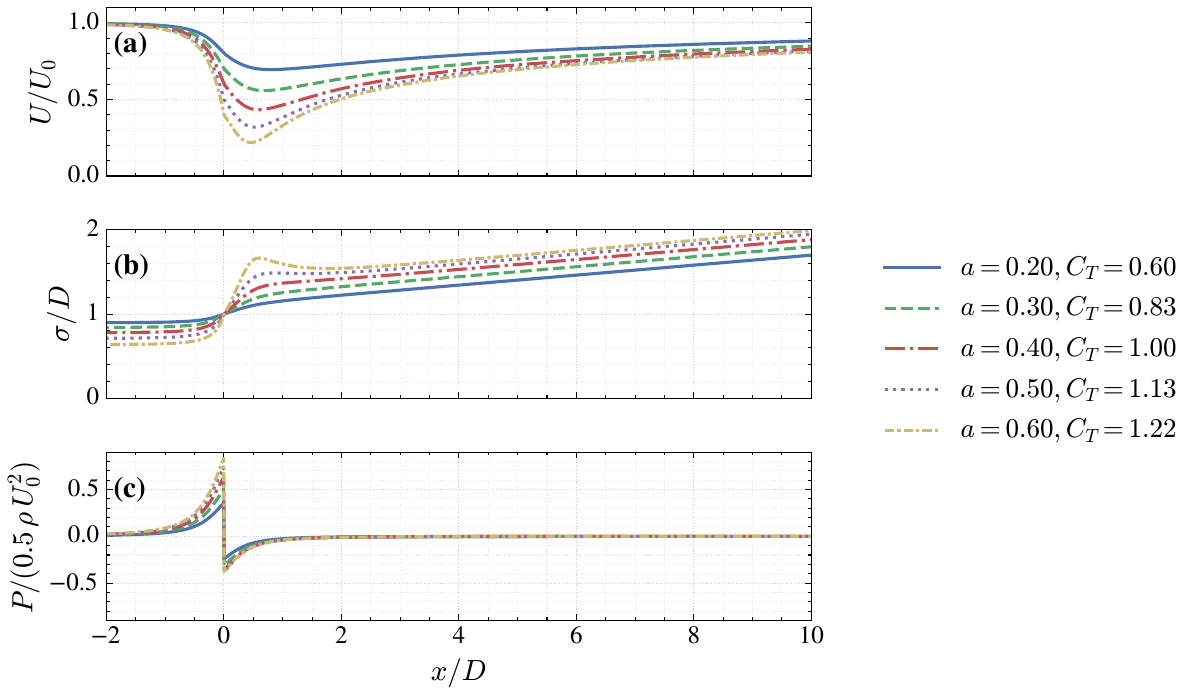}
    \caption{Streamwise variations of $U$, $\sigma$, and $P$ for different values of induction factor $a$. The entrainment coefficients are set to \( E_1 = 0.1 \) and \( E_2 = 0.6 \), and the incoming turbulence intensity $I=5\%$. A computational notebook is provided at \url{https://cocalc.com/share/public_paths/dd25c8045631c814897236f183abe7b40ec83ea7/Figure\%206}
 that computes $U(x/D)/U_0$, $\sigma(x/D)/D$ and $P(x/D)/\frac{1}{2} \rho U_0^2$ for given parameters $a$ (or $C_T$), $E_1$, $E_2$ and $I$.}
    \label{fig:different_a}
\end{figure}

Model predictions for five different values of the induction factor $a$ are shown in figure~\ref{fig:different_a}. For all cases, the entrainment coefficients are set to $E_1 = 0.1$ and $E_2 = 0.3$, and the incoming turbulence intensity is $I = 5\%$.
Figure~\ref{fig:different_a}a shows the streamwise variation of velocity. In all cases, as expected, the velocity upstream decreases from $U_0$ to $U_0(1-a)$ at the disk. Immediately downstream, it drops further, mainly due to pressure recovery ($\upd p/\upd x > 0$ resulting in $\upd U/\upd x<0$ according to equation \eqref{eq:dudx}), until reaching a minimum value (i.e., maximum deficit) approximately one rotor diameter downstream. Our model predicts the maximum deficit to be smaller than $2aU_0$ predicted by Froude’s theory, due to the inclusion of turbulent entrainment already beginning at $x=0$. Further downstream, when pressure is mostly recovered, flow entrainment becomes the dominant mechanism, resulting in $\upd U/\upd x > 0$ according to equation \eqref{eq:dudx}. Moreover, as expected, higher values of $a$ produce greater velocity deficits.

As shown in figure~\ref{fig:different_a}b, the cross-sectional width $\sigma$ of the CV generally increases with $x$ in both the upwind and downwind regions. According to equation \eqref{eq:dsigmadx}, velocity reduction ($\upd U/\upd x < 0$) contributes to cross-sectional expansion both upstream and downstream, with downstream growth further enhanced by flow entrainment. Figure ~\ref{fig:different_a}b also shows that $\sigma$ increases with $a$ downstream, indicating a wider wake for actuator disks with higher loading. In far-wake engineering models (for instance, \cite{bastankhah2014new}), this impact of disk loading on the wake width is often incorporated through an semi-empirical initial wake width that depends on the thrust coefficient $C_T$.  

An interesting feature is observed for $a = 0.5$ and $a = 0.6$, where the wake undergoes a rapid initial expansion followed by a contraction and then gradually re-expands due to turbulent entrainment. Such non-monotonic behaviour has been extensively reported in the literature (\cite{wilson1974,hansen2015aerodynamics,martinez-tossas_numerical_2022}, among others) and is often attributed to the formation of complex turbulent wake structures, including vortex rings, characteristic of the turbulent state with $a>0.4$. It is interesting that our simpler model reproduces such features. Nonetheless, we should bear in mind that the developed pressure relation, equation \eqref{eq:PPE_solution_semiinf_f=g=0}, is an approximate solution of the pressure Poisson equation for lightly-loaded disks. Therefore, model predictions for large values of $a$ should be interpreted with caution. Finally, figure~\ref{fig:different_a}c shows the pressure distribution, where the asymmetric pressure drop on the disk plane, discussed earlier in section \ref{sec:pressure_variation}, is evident, particularly for large values of $a$. The figure suggests that pressure is recovered downstream within the first two rotor diameters. This behaviour may however differ from some findings in the literature on highly-loaded disks \citep{martinez-tossas_numerical_2022,bempedelis_analytical_2022}, where pressure persists longer. As noted previously, this is likely because the simplified pressure relation used here is a first-order approximation for lightly-loaded disks, in which we neglected non-linear terms in the pressure Poisson equation, as discussed in details in Appendix \ref{sec:appendix}. For highly-loaded actuator disks, \cite{bempedelis_analytical_2022} developed a relation that describes the far-wake pressure as a function of the wake velocity deficit and entrainment velocity. While accurately predicting the wake pressure for highly-loaded disks is beyond the scope of the present study, future work could adopt an approach similar to that of \cite{bempedelis_analytical_2022} to model the non-linear terms in the pressure Poisson equation (terms $f_1$ and $f_2$, defined in \eqref{eq:f(x,r)} and \eqref{eq:g(x,r)}, respectively) as a function of the wake velocity deficit and entrainment velocity.
  
\subsection{Far-wake solution}\label{sec:far-wake_behaviour}
Here, we examine model predictions for the asymptotic far-wake case, where $x \to \infty$. In this case, both the normal and lateral pressure force terms in the momentum equation \eqref{eq:momentum} vanish, and $U\to U_0$. From the momentum equation \eqref{eq:momentum} for large values of $x$, we obtain
 \begin{equation}\label{infinite_mom_equation}
   \mathrm{d} \left(\sigma^{2} U  (U_{0}-U) \right)\approx \mathrm{d} \left(\sigma^{2}  (U_{0}-U)\right)\approx 0.
 \end{equation}
This means that
\begin{equation}\label{eq:sigma_asymp_intermediate}
    U_0\mathrm{d}\sigma^{2}\approx \textrm{d}\left(\sigma^2 U\right).
\end{equation}
Inserting equation \eqref{eq:sigma_asymp_intermediate} in the conservation of mass expression in \eqref{eq:sigma2u_new} leads to  
\begin{equation}\label{eq:sigma_far_wake}
    \frac{\upd \sigma}{\upd x}= 2\frac{U_e}{U_0}\quad \mathrm{if}\; x\to\infty.
\end{equation}

\subsubsection{Entrainment driven by wake shear} 
\begin{figure}
    \centering
    \includegraphics[width=1\linewidth]{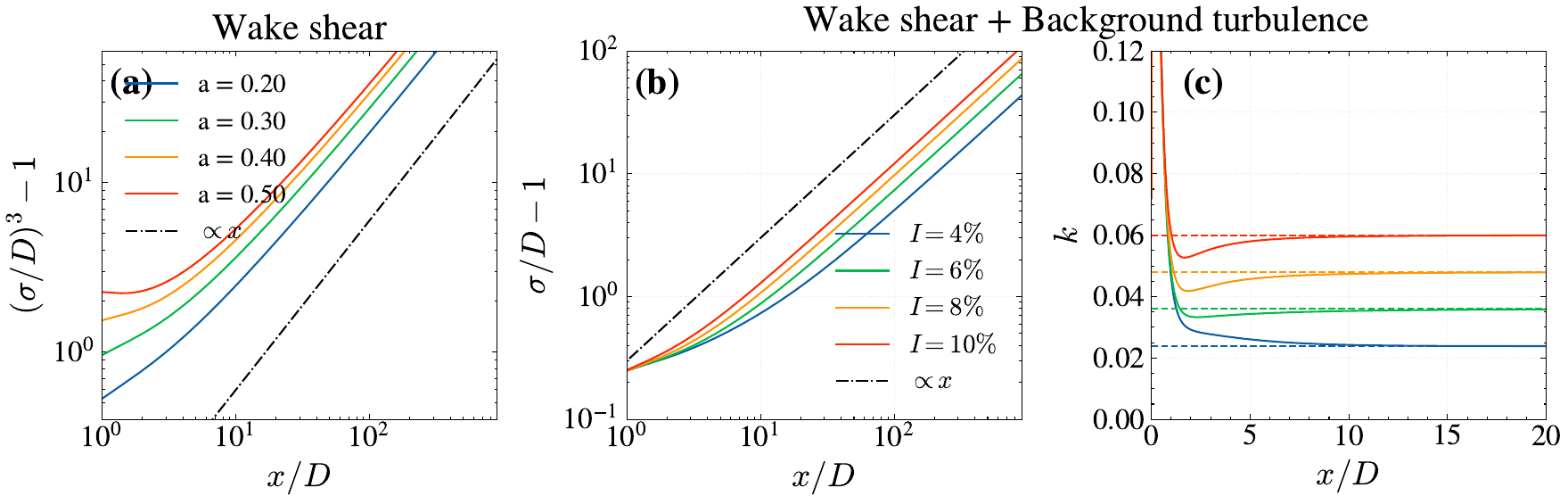}
    \caption{Asymptotic far-wake behaviour. (a) variations of $\sigma$ with $x$ where $E_1=0.1$ and $E_2=0$ for different values of induction factor $a$. (b) Variations of $\sigma$ with $x$ where $E_1=0.1$ and $E_2=0.6$ for $a=0.3$ and different values of incoming turbulence intensity $I$. (c) For the same dataset as in (b): variations of $k = 0.5\upd \sigma / \upd x$ with $x$ are shown as solid lines, with the asymptotic value $k_{\infty}$ included as a dashed line. }
    \label{fig:far_wake}
\end{figure}

First, we assume the ambient turbulence is negligible, and the main driving force for turbulent entrainment is the wake shear. In this case,  $E_2=0$, so $U_e=E_1(U_0-U)$. From the simplified form of the momentum equation \eqref{eq:momentum} for $x\to \infty$ (i.e., no pressure effects and $U\to U_0$), we substitute $U_0-U$ to have
\begin{equation}\label{eq:U_e:case1}
    U_e=E_1(U_0-U)=\left(\frac{1}{2}E_1C_TU_0D^2\right)\sigma^{-2}.
\end{equation}
Inserting equation \eqref{eq:U_e:case1} in equation \eqref{eq:sigma_far_wake} and integrating leads to
\begin{equation}\label{eq:sigma_inf_case1}
    \frac{\sigma}{D}=\left(3E_1C_T\right)^{1/3}\left(\frac{x-x_0}{D}\right)^{1/3},
\end{equation}
where $x_0<0$ is the virtual origin defined as the location where $\sigma(x_0)=0$. There are two interesting points about equation \eqref{eq:sigma_inf_case1}. First, it shows that the far wake grows with downstream distance as $x^{1/3}$ in the absence of ambient turbulence, as reported in previous studies that assumed $U_e\propto (U_0-U)$ \citep{luzzatto-fegiz_one-parameter_2018}. This is consistent with the extensive literature of three-dimensional wakes subject to laminar free-stream flows (see \citet{pope2001turbulent} and references therein). Moreover, this shows that the spreading rate of the wake depends on $C_T$. This supports George's theory \citep{george2004role,johansson2003equilibrium} that although canonical free shear flows may reach universal asymptotic states in terms of self-similarity, their spreading rate is not universal, and it depends on upstream (i.e., initial) conditions. This is clear in figure \ref{fig:far_wake}(a), which shows variations of $\sigma^3$ with respect to $x$ for different values of $a$, where $E_1=0.1$ and $E_2=0$. To verify equation~\eqref{eq:sigma_inf_case1}, 
we rearrange it assuming that \(x_0\) is chosen such that \(\sigma(x=0)=D\). 
Under this assumption, equation~\eqref{eq:sigma_inf_case1} simplifies to 
\((\sigma/D)^3 - 1 \propto x\), 
which further reduces to \((\sigma/D)^3 \propto x\) when \(\sigma \gg D\), as shown in figure \ref{fig:far_wake}(a).

\subsubsection{Entrainment driven by both wake shear and background turbulence}\label{sec:far_wake_ambient}
For the case where large-scale background turbulence is present, equation \eqref{eq:entrainment_velocity} simplifies to $U_e/U_0 = E_2 I$ as $x \to \infty$. Substituting this relation into equation \eqref{eq:sigma_far_wake} and integrating yields
\begin{equation}\label{eq:sigma_inf_case2}
\frac{\sigma}{D} = 2k_\infty \left(\frac{x - x_0}{D}\right),
\end{equation}
where $k_{\infty} = \frac{1}{2} \upd  \sigma_{\infty} / \upd x = E_2 I$ is the expansion rate of the far-wake radius. The parameter $k_{\infty}$ is an empirical constant which is also widely used in far-wake engineering models of wind turbines \citep[e.g., Park model developed by][]{Jensen1983, Katic1986}. \citet{pena_application_2016} suggested $k_{\infty} = 0.4 I$, which corresponds to $E_2 = 0.4$. Traditionally, values of $k_{\infty} = 0.04-0.05$ are used for offshore cases \citep{Barthelmie2009ModellingOffshore,barthelmie2010}. More recently, considering wind farm wakes and wake evolution at large downwind distance, \cite{nygaard_large-scale_2022} reported that $k_{\infty} = 0.02$ provides results in closer agreement with operational data from offshore wind farms. For offshore turbulence intensity at turbine hub height in the range of $5\%$–$7.5\%$, these reported values of $k_{\infty}$ translates to $E_2 = 0.25$–$1$. While the aim of our study is not to determine the universal value of $E_2$, the validation analysis presented in section \ref{sec:validation} against multiple numerical and experimental datasets suggest $E_2=0.6$. Variations of $\sigma/D-1$ with $x/D$ are shown in figure \ref{fig:far_wake}(b) for different values of incoming turbulence intensity $I$. In this figure, $E_1 = 0.1$, $E_2 = 0.6$, and $a = 0.3$ are used. Similar to the approach used in figure~\ref{fig:far_wake}(a) and to verify equation~\eqref{eq:sigma_inf_case2}, here we plot \(\sigma/D - 1\), which is basically the normalised increase of the wake width with respect to its initial value.
 The results illustrate  that linear wake growth begins earlier for higher values of incoming turbulence intensity. For the same parameter set, the corresponding values of $k(x) = \frac{1}{2} \upd   \sigma / \upd x$ are shown in figure \ref{fig:far_wake}(c) as solid lines, with the asymptotic value $k_{\infty}$ included as a dashed line for reference. This figure shows that the wake expansion rate $k$ asymptotes to $k_{\infty}$ at large downwind distances. 
This variable wake recovery rate is consistent with observations in the literature \citep{nygaard2020modelling,vahidi2022physics}.

\subsubsection{Far-wake behaviour for a generic entrainment velocity relation}\label{section:far_wake_generic}
In general, if the entrainment velocity in the far wake is expressed as $U_e \propto (U_0-U)^{\alpha}u_*^{1-\alpha}$, where $\alpha$ is a constant and $u_*$ is the characteristic velocity scale of the background turbulence (e.g., the root-mean-square of velocity fluctuations or the friction velocity), one can conclude from \eqref{infinite_mom_equation} and \eqref{eq:sigma_far_wake} that $\sigma \propto x^{1/(1+2\alpha)}$ and $(U_0-U) \propto x^{-2/(1+2\alpha)}$. For $\alpha=1$ and $\alpha=0$, these expressions simplify to the well-known scalings of $\sigma \propto x^{1/3}$ and $\sigma \propto x$, respectively, as discussed earlier. 

Other scalings can, however, be retrieved by choosing a different value for $\alpha$. For instance, several studies have suggested that an axisymmetric turbulent wake may scale with $\sigma \propto x^{1/2}$. \cite{nedic_axisymmetric_2013} made this prediction based on the assumption of nonequilibrium turbulence dissipation. \cite{eames2011Johnson_sphere} suggested that $\sigma \propto x^{1/2}$ may occur in the very far-field wake, where $(U_0-U) \ll u_*$ and $\sigma \gg \Lambda$. In this case, according to Taylor's dispersion theory \citep{taylor1922diffusion}, the turbulent eddies are much smaller than the wake length scale, such that they effectively act like molecular diffusion and the wake evolves with a constant diffusivity coefficient (akin to laminar wakes). Similar conclusion was drawn by \cite{johansson2003equilibrium} based on a different approach. They showed that the local Reynolds number ($\text{Re}_l=\sigma (U_0-U)/\nu$, where $\nu$ is the kinematic viscosity) of an axisymmetric wake diminishes with streamwise distance. This ultimately leads to the wake transitioning from a high to a low local Reynolds number regime, where it spreads as $\sigma \propto x^{1/2}$. The choice of $\alpha=0.5$ for the entrainment velocity (i.e. $U_e^2\propto (U_0-U)u_*$) leads to this scaling.

\subsection{Comparison of flow predictions with experimental and numerical data}\label{sec:validation}
In this section, we compare flow predictions based on the proposed actuator disk theory with three different datasets, covering both LES and wind-tunnel measurements across a range of inflow conditions and $C_T$ values. For all comparisons shown in this section, we use a wake-shear entrainment coefficient of $E_1=0.1$ \citep{morton_momentum-mass_1961}, and a background-turbulence entrainment coefficient of $E_2=0.6$.

\subsubsection{LES data of a porous disk in free-stream turbulence}
We compare predictions of the proposed actuator disk model against the LES data of \cite{li_impacts_2024}. The reference study investigated the flow past a porous disk subject to uniform inflow with synthetic turbulence \citep{mann1994spatial}, covering a range of thrust coefficients, turbulence intensities, and integral length scales. From these datasets, we selected the inflow with an integral length scale of $\Lambda = 1.5D$ to best approximate typical turbine operating conditions in the atmosphere. The comparison is performed for two thrust coefficients ($C_T = 0.2$ and $0.7$) and two ambient turbulence intensities ($I = 10\%$ and $25\%$). The LES data are reported as averages over the fixed disk area, which may introduce discrepancies when they are compared with our model predictions, particularly in the far wake where the physical wake expands beyond the rotor disk area.
\begin{figure}
    \centering
    \begin{overpic}[width=1\linewidth]{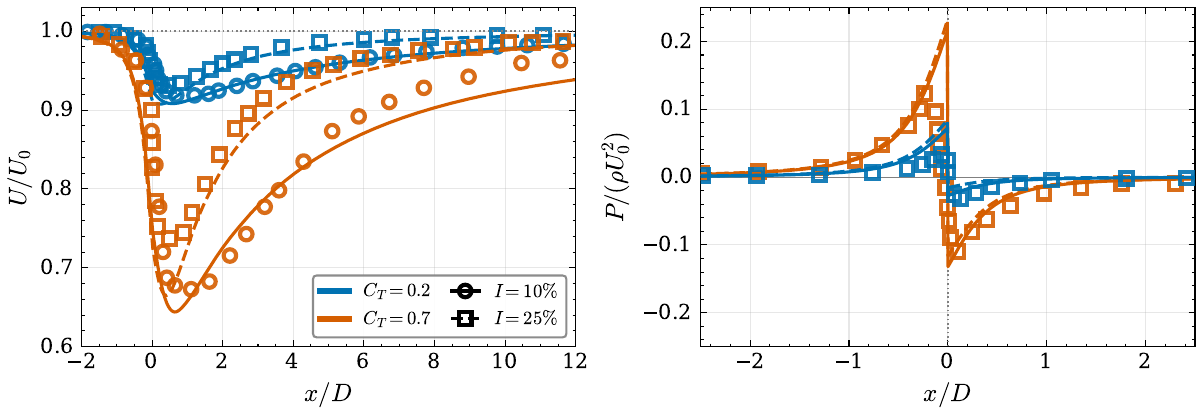}
        \put(-5,120){{(a)}} 
        \put(195,120){{(b)}}
    \end{overpic}
    \caption{Comparison of actuator-disk model predictions (lines) against LES data of \cite{li_impacts_2024} (markers) for (a) normalised velocity $U/U_0$ and (b) normalised pressure $P/(\rho U_0^2)$. Results are presented for two thrust coefficients, $C_T=0.2$ (blue) and $C_T=0.7$ (orange), at ambient turbulence intensities of $I=10\%$ (LES: circles, Model: solid lines) and $I=25\%$ (LES: squares, Model: dashed lines).}
    \label{fig:Lie_LES}
\end{figure}

Figure \ref{fig:Lie_LES} compares the model predictions with the LES data. In terms of velocity (figure~\ref{fig:Lie_LES}a), the model captures the expected physical trends: the wake recovers faster with increased ambient turbulence intensity ($I$), and the velocity deficit is more pronounced for the higher thrust coefficient ($C_T=0.7$). Overall, the agreement between the model and LES is good for all four cases considered. Regarding the pressure distribution (figure~\ref{fig:Lie_LES}b), the model tends to overpredict the magnitude of the pressure increase upstream of the disk. This discrepancy likely arises because the pressure value used in the model is the centreline pressure computed from \eqref{eq:P_D}, whereas the LES data represent disk-averaged values. Furthermore, the disk-averaged pressure values obtained from LES can be sensitive to the force smoothing regularisation applied at the edges of the actuator disk. It is worth noting that the LES study reported no noticeable impact of inflow turbulence on pressure variations; consequently, LES pressure data are plotted for only one inflow condition in figure ~\ref{fig:Lie_LES}b. The model predicts that inflow turbulence has a small effect on the pressure field. A more detailed discussion on the impact of inflow turbulence on pressure and rotor performance is provided in section \ref{sec:ct_vp}.

\subsubsection{Wind-tunnel data of a porous disk in free-stream turbulence}\label{sec:wind_tunnel_validation}
\begin{figure}
    \centering
    \includegraphics[width=1\linewidth]{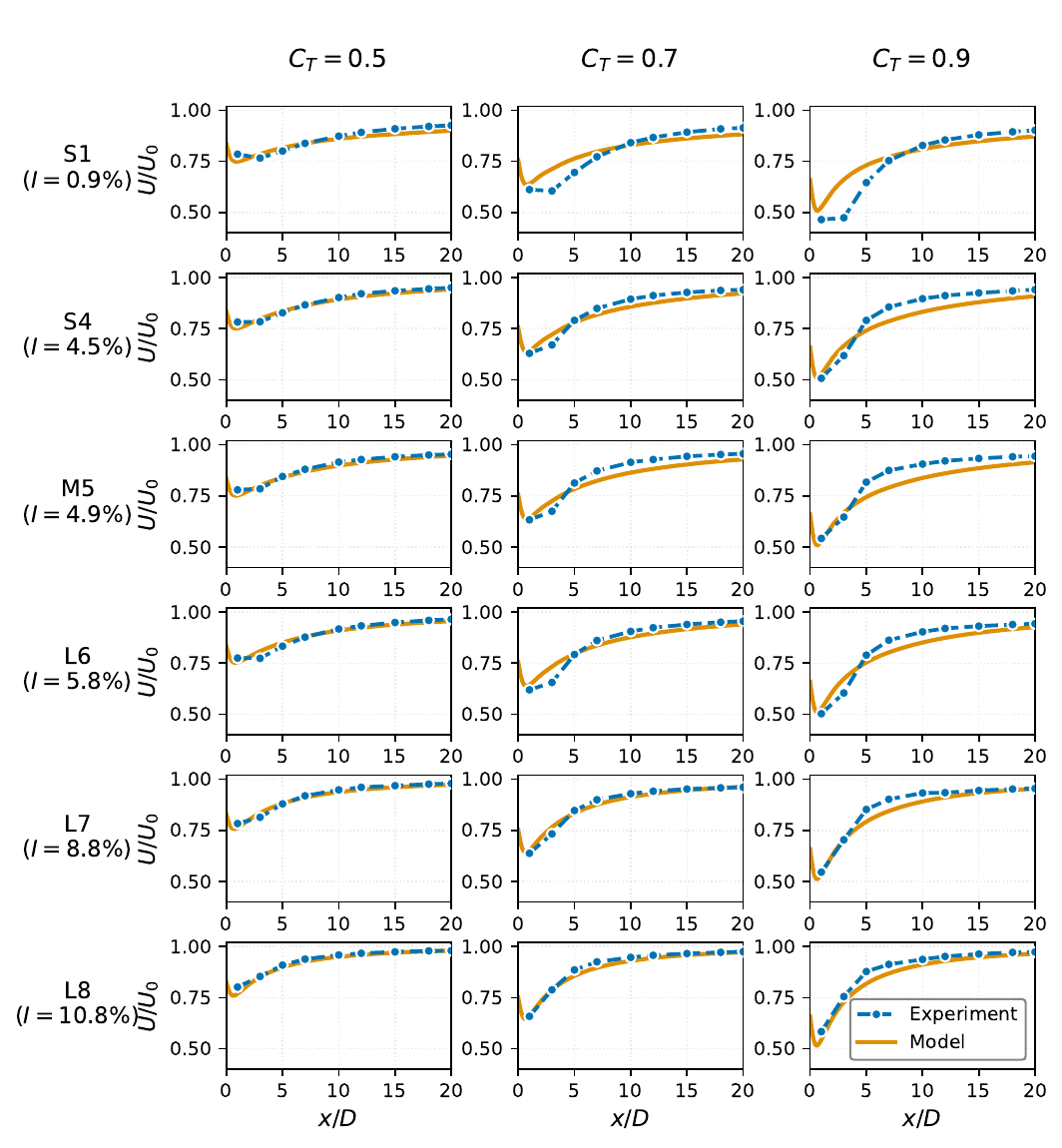}
\caption{Streamwise evolution of the normalized velocity $U/U_0$. The comparison shows model predictions (orange solid lines) against the experimental data of \cite{bourhis_impact_2025} (blue markers). The panels are organised in a grid where columns represent different thrust coefficients ($C_T \in \{0.5, 0.7, 0.9\}$) and rows represent different inflow conditions with increasing ambient turbulence intensity ($I$) ranging from $0.9\%$ (S1) to $10.8\%$ (L8). The inflow labels follow the nomenclature used in the original study.}
    \label{fig:bourhis_velocity}
\end{figure}
Here, we compare model predictions against the wind tunnel measurements of \cite{bourhis_impact_2025}, which characterise the wake of a porous disk subjected to grid-generated free-stream turbulence. We use a portion of this dataset that covers a broad parameter space, comprising three thrust coefficients ($C_T \in \{0.5, 0.7, 0.9\}$) and six distinct inflow conditions with turbulence intensities ranging from $I \approx 0.9\%$ to $10.8\%$, amounting to 18 different cases. To enable a direct comparison with the top-hat profiles predicted by the actuator disk model, the experimental transverse velocity profiles were projected onto equivalent top-hat shapes defined by a uniform wake velocity $U$ and a wake width $\sigma$. These two unknowns are determined uniquely by equating the mass flux deficit ($m$) and momentum flux deficit ($M$) of the experimental profile to those of the equivalent top-hat profile. Specifically, if $U_{\rm exp}(x,r)$ is the measured velocity and $dA = 2\pi r dr$ is the elemental area, we calculate $
    m(x) = \int \left(U_0 - U_{\rm exp}(x,r)\right) dA,$ and $M(x) = \int U_{\rm exp}(x,r)\left(U_0 - U_{\rm exp}(x,r)\right) dA.$
The equivalent uniform velocity $U$ and width $\sigma$ are then derived algebraically from the conservation of these two quantities:
\begin{equation}
    U = \frac{M}{m}, \quad \sigma = \sqrt{\frac{4}{\pi} \frac{m^2}{U_0 m - M}}.
\end{equation}
An advantage of this flux-based approach is that it forms a closed system for the two unknowns ($U, \sigma$) without relying on arbitrary velocity thresholds (e.g., defining the wake edge at 10\% wake-centre deficit).

\begin{figure}
    \centering
    \includegraphics[width=1\linewidth]{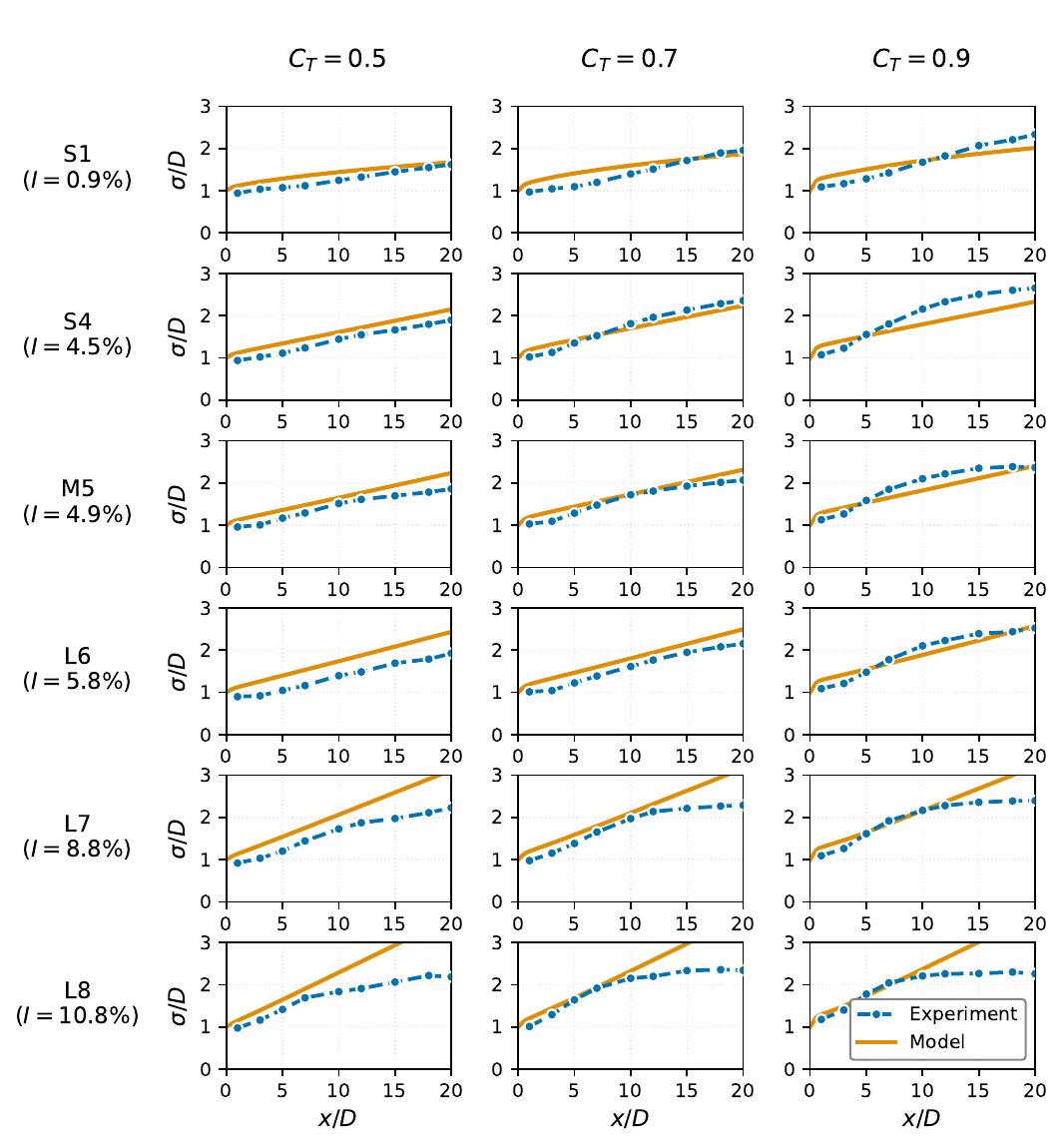}
\caption{Streamwise evolution of the normalized wake width $\sigma/D$. The comparison shows model predictions (orange solid lines) against the experimental data of \cite{bourhis_impact_2025} (blue markers). The figure layout follows the same convention as figure~\ref{fig:bourhis_velocity}.}
    \label{fig:bourhis_sigma}
\end{figure}

Figure \ref{fig:bourhis_velocity} presents a comparison of the streamwise velocity evolution downstream of the disk between the model predictions and the wind tunnel measurements. Note that the labelling convention for the inflow conditions (e.g., S1, L8) follows the nomenclature used in the original study. The figure shows that overall velocity predictions are in agreement with the experimental data across the majority of the parameter space. However, a deviation is observed for cases combining high thrust coefficients ($C_T=0.9$) with low ambient turbulence intensities (e.g., case S1 with $C_T=0.9$). In this regime, the model tends to overpredict the near-wake velocity immediately downstream of the disk.

The comparison for the normalised wake width evolution $\sigma/D$ is shown in Figure~\ref{fig:bourhis_sigma}. While the agreement is fairly satisfactory in the near-to-intermediate wake ($x < 10D$), discrepancies emerge in the far wake ($x > 10D$), particularly for cases with high turbulence intensity (e.g., L7 and L8). In these instances, the model predicts a continued expansion of the wake, whereas the experimental data indicate a distinct reduction in the spreading rate. This difference is likely due to the formulation of the background-turbulence entrainment term in equation~\eqref{eq:entrainment-atmospheric}. This formulation assumes a constant entrainment velocity based on the inflow turbulence level, leading to a linear wake expansion, as shown in section~\ref{sec:far-wake_behaviour}. This assumption is expected to be valid for actuator disks subjected to large-scale atmospheric eddies, which will be verified later in section \ref{section:validation-LES}. However, in laboratory experiments with uniform inflows where free-stream turbulence has small or moderate integral length scales ($\Lambda \leq 2D$ for this dataset), the entrainment velocity requires further refinement to accurately capture the asymptotic wake expansion. See the discussion in Section \ref{section:far_wake_generic} for more details on the impact of the entrainment velocity model on wake expansion.

\subsubsection{LES data of a turbine in atmospheric turbulence}\label{section:validation-LES}
\begin{figure}
    \centering
    \includegraphics[width=.7\linewidth]{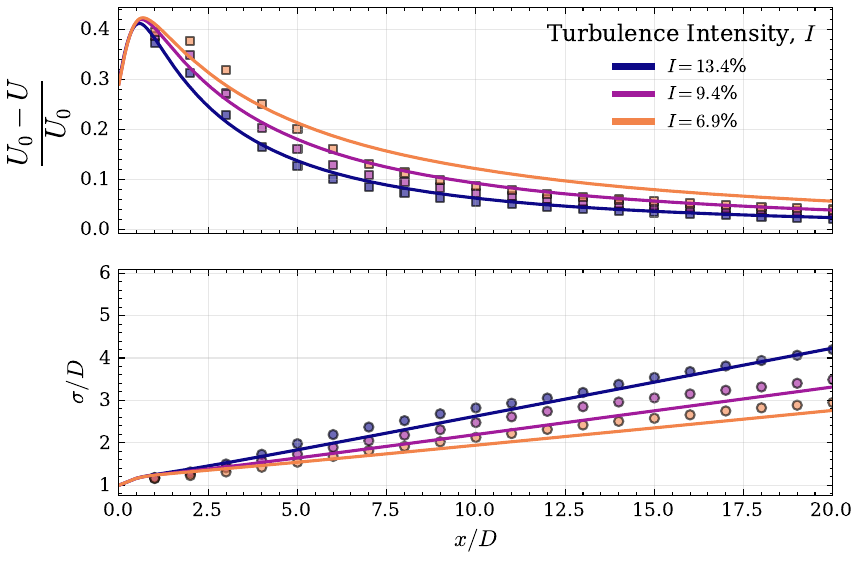}
\caption{Comparison of model predictions (solid lines) against the LES data of \cite{Wu2012} (markers) for a wind turbine operating at $C_T=0.8$. The panels show the streamwise evolution of (a) the normalised velocity deficit $\Delta U / U_0$ and (b) the normalised wake width $\sigma/D$. Results are presented for three ambient turbulence intensities: $I=6.9\%$, $9.4\%$, and $13.4\%$. The experimental wake characteristics $U$ and $\sigma$ are derived using the same flux-conservation approach used for the wind-tunnel data in section~\ref{sec:wind_tunnel_validation}.}
\label{fig:Wu_LES}
\end{figure}

Finally, we compare model predictions against the LES data of \cite{Wu2012}, which simulate a utility-scale wind turbine operating in a neutral atmospheric boundary layer. The turbine operates at a thrust coefficient of $C_T=0.8$ and is subjected to three different inflow conditions with hub-height turbulence intensities of $I=6.9\%$, $9.4\%$, and $13.4\%$. Consistent with the methodology applied to the wind tunnel measurements, the characteristic wake velocity $U$ and wake width $\sigma$ were extracted from the LES flow fields using the flux-conservation approach described in Section~\ref{sec:wind_tunnel_validation}. Since the wake expands asymmetrically in the atmospheric boundary layer (due to shear and ground effects), the reported values for $U$ and $\sigma$ represent the average of the characteristics derived from the horizontal ($xy$) and vertical ($xz$) mid-planes. The comparison between the model predictions and the LES data is presented in figure~\ref{fig:Wu_LES}. Overall, the model demonstrates good agreement with the simulation results for both the velocity deficit and the wake expansion rate across the tested turbulence intensities. The most notable discrepancy is observed in the far wake ($x > 10D$) for the lowest turbulence intensity case ($I=6.9\%$), where the model underpredicts the velocity recovery. 

\subsection{Relation between thrust coefficient and induction factor}\label{sec:ct_vp}

\begin{figure}
    \centering
    \includegraphics[width=1\linewidth]{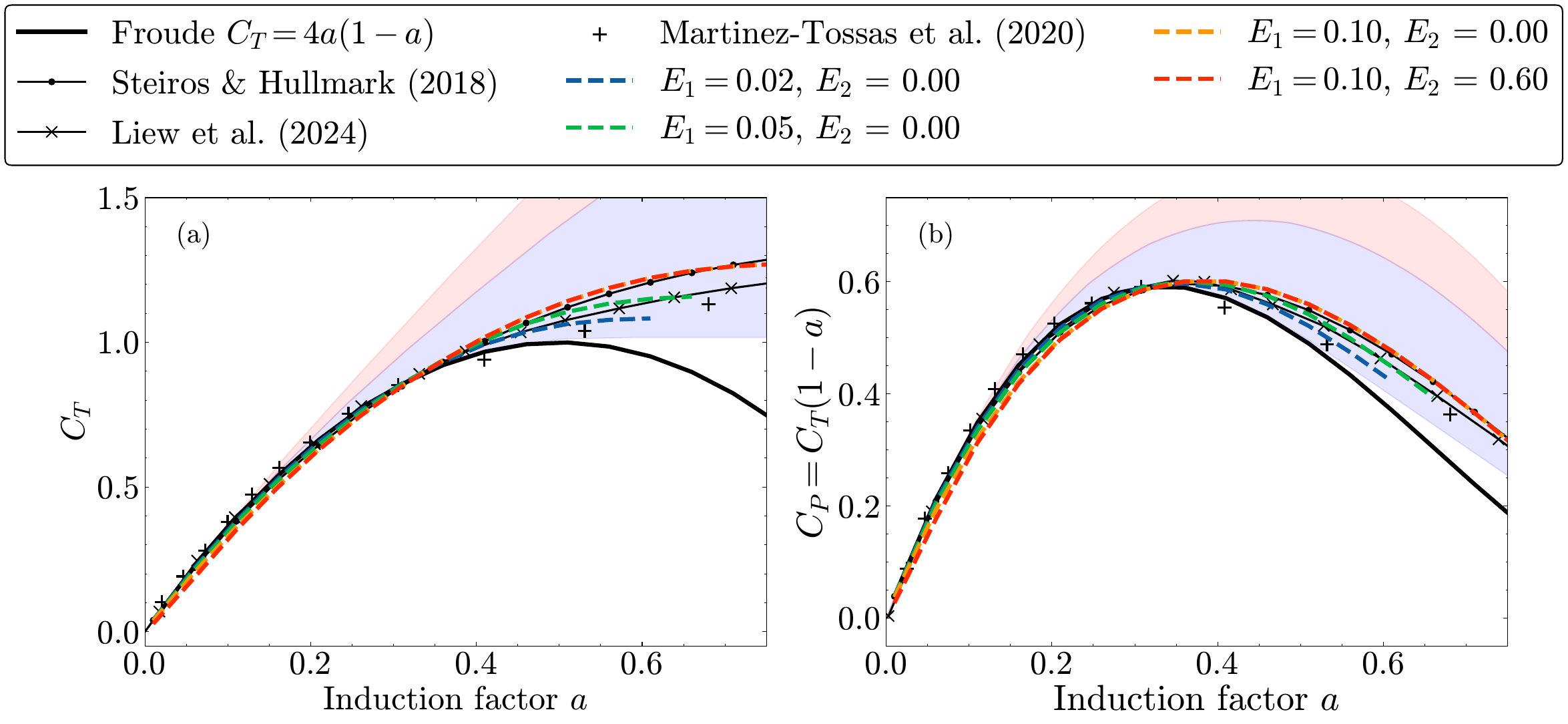}
    \caption{Variations of thrust coefficient $C_T$ (panel a) and power coefficient $C_P$ (panel b) for different values of entrainment coefficients $E_1$ and $E_2$, where the incoming turbulence intensity $I=5\%$. As reference, predictions of Froude's actuator disk theory, more recent models of \citet{steiros_drag_2018} and \citet{liew_unified_2024}, and the LES data of \cite{martinez-tossas_numerical_2022} (constant $C_T$ method) are also shown. Additionally, the theoretical limits from the model of \citet{Dehtyriov2023} are shown as shaded regions: the blue shaded area represents the operating zone between the lower mixing bound and the far-wake mixing limit, while the red shaded area represents the zone between the far-wake mixing limit and the near-wake mixing limit. A computational notebook is provided at \url{https://cocalc.com/share/public_paths/dd25c8045631c814897236f183abe7b40ec83ea7/Figure\%2012} to compute $C_T$ and $C_P$ for given parameters $a$, $E_1$, $E_2$, and $I$.}
    \label{fig:ct_cp}
\end{figure}

Next, we discuss the predictions of thrust coefficient $C_T$ and power coefficient $C_P = C_T (1-a)$ based on the developed actuator disk model and the iterative process explained in Section \ref{sec:determine ct}. Figure~\ref{fig:ct_cp} shows variations of $C_T$ and $C_P$ with the induction factor $a$ for different values of $E_1 = (0.02, 0.05, 0.1)$ and $E_2 = (0, 0.6)$. For comparison, the figure also shows Froude's relation ($C_T=4a(1-a)$), LES data of \cite{martinez-tossas_numerical_2022}, and the models of \cite{steiros_drag_2018} and \cite{liew_unified_2024}. The figure shows that in the windmill state ($a < 0.4$), model predictions are not very sensitive to the value of entrainment coefficients and provide results similar to Froude's theory, albeit with slightly smaller values. On the other hand, in the turbulent state ($a > 0.4$), results are highly sensitive to the value of $E_1$, and the inclusion of entrainment considerably improves predictions of actuator disk models in comparison to the unphysical predictions of Froude's theory in this region, where higher values of $E_1$ lead to higher values of $C_T$. A value of $E_1 = 0.05$ seems to provide results in satisfactory agreement with the model of \cite{liew_unified_2024}, validated against LES data, while a value of $E_1 = 0.1$ provides results in very good agreement with the model of \cite{steiros_drag_2018}, validated against laboratory experiments. It is worth noting that the iterative method described in Section \ref{sec:determine ct} does not converge for cases with high values of $a$ (e.g., $a > 0.6$) and small values of $E_1$ (e.g., $E_1 = 0.02$). This may be due to the fact that, in highly loaded cases, turbulent mixing and entrainment play such an important role in shaping the wake structure downstream of the disk that minimising their effect does not lead to meaningful  predictions. Furthermore, while the linearised pressure relation in equation \eqref{eq:PPE_solution_semiinf_f=g=0} is fundamentally a first-order approximation, figure \ref{fig:ct_cp} shows that the model still provides realistic predictions of $C_T$ and $C_P$ beyond classical Froude theory's limit of $a \sim 0.4$. As discussed in section \ref{sec:pressure_variation}, this success is driven not only by the inclusion of turbulent entrainment but also by directly enforcing the exact pressure jumps derived from the Bernoulli equation as boundary conditions in the partial differential equation for pressure 
solution. This approach reduces the errors that would otherwise arise from neglecting non-linear terms in highly-loaded regimes. 

In reality, the entrainment coefficient $E_1$ may not be strictly constant and may have some dependencies on the actuator disk characteristics, inflow properties and streamwise location (i.e., $E_1 = E_1(x/D,a,I)$). However, the key physical insight provided by figure \ref{fig:ct_cp} is that a universal relation between $a$ and $C_T$ may not exist; rather, their exact relationship depends on the interaction of the actuator disk flow with the ambient flow and the turbulent entrainment, particularly immediately downstream of the disk. This is especially important for highly-loaded disks, where the interaction between the wake and the surrounding flow is strong and cannot be neglected. This dependence may also explain why observations of $C_T = C_T(a)$ performed in different studies typically show good agreement in the windmill state, while the data become widely scattered in the turbulent state, as shown in figure \ref{fig:ct_cp} and reported in previous works \citep[e.g.][]{buhl2005new}.

Figure \ref{fig:ct_cp} also shows the theoretical limits derived by \cite{Dehtyriov2023} for $C_T$ and $C_P$ of an actuator disk that is subject to turbulent wake mixing. \cite{Dehtyriov2023} considered the region behind an actuator disk as a sequence of small control volumes in which the core wake flow mixes with the ambient flow, and this mixing continues until the wake velocity is fully recovered. This led to determining three theoretical limits (i.e., bounds) based on far-wake mixing and near-wake mixing. Building on the work of \cite{nishino_efficiency_2013}, \cite{Dehtyriov2023} defined near-wake mixing as the mixing that occurs right behind the actuator disk where the pressure in the core of the wake is different from the surrounding layer (i.e., bypass flow). Far-wake mixing, on the other hand, is defined as the mixing that occurs after the pressure in the core of the wake is equal to the pressure of the surrounding layer. The bottom of the blue shaded region in Figure \ref{fig:ct_cp} shows model predictions derived by modelling far-wake mixing as a sudden single-step process. The upper limit for gradual (i.e., sequential) far-wake mixing is shown as the top of the blue shaded region in the figure, and finally, the upper limit of the near-wake mixing is shown as the top of the red shaded region. The figure shows that experimental and LES data, as well as other model predictions, lie between the sudden and sequential far-wake mixing theoretical limits.

Finally, the figure indicates that the value of \( E_2 \) has minimal impact on the variation of \( C_T \) and \( C_P \). This is expected, as \( E_2 \) mainly governs entrainment in the far-wake region where background turbulence effects dominate, whereas the near-wake region, where wake-shear driven entrainment mainly occurs, has a more direct influence on the thrust force and power output. Therefore, the model predicts that background turbulence has a negligible impact on both $C_T$ and $C_P$, which is consistent with the experimental findings of \cite{graham_turbulent_1976}. Background-turbulence driven entrainment is expected to have some possible impact on $C_T$ only in the asymptotic case of $a \to 0$, since in this case the entrainment induced by ambient turbulence can dominate even in the region immediately behind the turbine.  It is also worth noting that our model does not account for the increase in total kinetic energy (i.e., mean plus turbulent components) of the incoming flow caused by background turbulence. Because power production has a non-linear relationship with incoming velocity, this additional turbulent kinetic energy may increase power output, as documented in previous studies \citep[e.g.,][]{sheinman_dynamic_1992, gambuzza2021effects}.


\subsection{Optimal power and revised Betz limit}

\begin{figure}
    \centering
    \includegraphics[width=.9\linewidth]{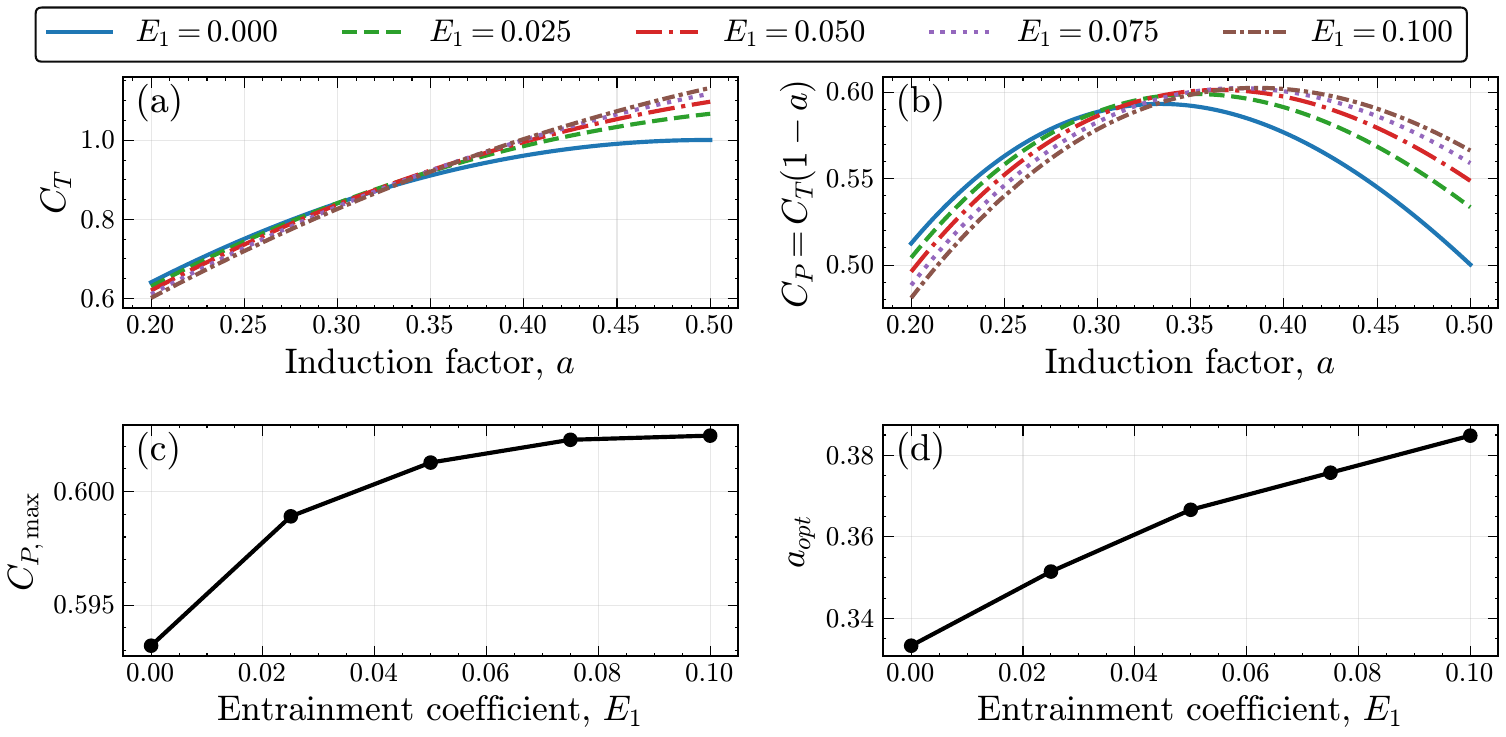}
   \caption{Zoomed-in view of the (a) $C_T$ and (b) $C_P$ versus induction factor $a$ near the maximum-$C_P$ operating point, for various values of the turbulent entrainment parameter $E_1$. Panels (c) and (d) show the maximum power coefficient $C_{P,max}$ and the corresponding optimal induction factor $a_{opt}$, respectively, as functions of $E_1$. Background-turbulence driven coefficient $E_2$ is set to zero in this figure.}
    \label{fig:cp-blowup}
\end{figure}

Here, we examine predictions of the new theory for the maximum power coefficient $C_{P,\max}$ and its corresponding optimal induction factor $a_{\text{opt}}$. Figures~\ref{fig:cp-blowup}(a) and (b) show zoomed-in portions of the $C_T$–$a$ and $C_P$–$a$ curves near $a_{\text{opt}}$, while figures~\ref{fig:cp-blowup}(c) and (d) illustrate the variations of $C_{P,\max}$ and $a_{\text{opt}}$ as functions of $E_1$. The results indicate that as the turbulence parameter $E_1$ increases, $a_{\text{opt}}$ exceeds its classical value of $1/3$, and $C_{P,\max}$ slightly surpasses the Betz limit of $16/27 \approx 0.593$. Exceeding the Betz limit can be explained by turbulent mixing, which enables a greater recovery of the negative gauge pressure ($P_D^-$) behind the disk to its ambient value ($P=0$). Therefore, in the presence of turbulence, the magnitude of $|P_D^-|$ can be larger than in the non-turbulent case. The thrust coefficient depends on the pressure drop across the disk ($\frac{1}{2}\rho U_0^2 C_T = P_D^+ - P_D^-$), and the upstream pressure ($P_D^+$) is unaffected by downstream mixing. Therefore, a larger $|P_D^-|$ leads to a larger $C_T$ and also a larger $C_P = C_T(1-a)$. 

Exceeding the Betz limit can be also explained by equation \eqref{eq:f_ps+=f_ps-}, which dictates that the integral of the product of wake expansion and pressure must remain constant. In Froude's classical actuator disk theory, the lack of wake recovery by turbulent entrainment causes the model to overpredict the initial flow expansion at high values of $a$, where the wake area approaches $\infty$ as $a \to 0.5$. To satisfy equation \eqref{eq:f_ps+=f_ps-}, this unrealistic flow expansion forces an unrealistic reduction in the predicted magnitude of $P_D^-$, which in turn lowers $C_T$ and $C_P$ at high induction factors. In our framework, however, turbulent mixing restrains the wake expansion to more moderate, realistic levels due to turbulent recovery. This is shown in figure ~\ref{fig:different_E}(b) for $a=0.45$ (comparing the mixing case, $E_1 > 0$, with the no-mixing case, $E_1=0$). This moderated flow expansion allows for a larger $|P_D^-|$ (and therefore larger $C_T$ and $C_P$), explaining why the classical limit can be exceeded. Values of $C_P$ exceeding the Betz limit have been reported in several previous studies \citep[e.g.,][]{nishino2012efficiency, Dehtyriov2023, liew_unified_2024} due to factors such as flow mixing and flow confinement.

It is worth noting that for $a < 1/3$, the effect of turbulent mixing is the opposite, as seen in figures~ \ref{fig:cp-blowup}(a) and (b). In this case, mixing slightly increases the near-wake expansion (see figure~\ref{fig:different_E}(a)), which reduces $|P_D^-|$ according to equation \eqref{eq:f_ps+=f_ps-} and consequently lowers both $C_T$ and $C_P$. As already shown, however, the influence of turbulence mixing on the rotor performance for $a < 1/3$ is much weaker than for $a > 1/3$. In reality, the impact of wake turbulent mixing on rotor performance for small induction factors ($a < 1/3$) may be even smaller than shown here if entrainment does not begin immediately behind the disk; for instance, if shedding tip vortices act as a barrier between the wake and the ambient flow \citep{lignarolo2015tip}. In the current formulation, we assume that wake-shear-driven entrainment \eqref{eq:entrainment_wake_shear} starts immediately behind the disk, while background-turbulence-driven entrainment \eqref{eq:entrainment-atmospheric} is zero at $x=0$ and reaches its maximum value further downstream.

\section{Summary and conclusions}
In this work, we develop a generalised actuator disk theory that extends the classical formulation by:
(i) describing how flow properties vary with the streamwise coordinate $x$, rather than limiting the analysis to far-upstream and far-downstream conditions;
(ii) incorporating the effects of turbulent mixing in the wake and its contribution to wake recovery; and
(iii) providing physically consistent predictions of the thrust coefficient $C_T$ for a given induction factor $a$ (or vice-versa) for highly-loaded disks.  

The mass and momentum equations are formulated for a hybrid CV encompassing the streamtube and the wake, upstream and downstream of the actuator disk, respectively. In the downwind region, wake recovery is modelled as the result of turbulent entrainment across the lateral surface of the CV with an entrainment velocity that depends on both the wake shear and the ambient turbulence. Moreover, it is shown that, at finite streamwise positions, pressure forces acting on the lateral surface of the CV contribute a non-negligible term to the axial momentum balance and must be included for a self-consistent formulation. 

By integrating mass and momentum balances between far upstream and an arbitrary downstream position $x$, we derive a generalised Bernoulli-type equation that accounts for both energy extraction by the disk and energy injection into the wake via turbulent mixing. An additional pressure equation is obtained by solving a simplified, axisymmetric form of the pressure Poisson equation, yielding accurate predictions for lightly-loaded disks when compared with LES data. Together, the system consists of three governing relations: two differential equations (obtained from mass and momentum equations) and one algebraic equation (obtained from an approximate solution of the pressure Poisson equation), which can be solved to obtain the streamwise evolution of velocity $U(x)$, pressure $P(x)$, and CV diameter $\sigma(x)$ for given values of the induction factor $a$ and thrust coefficient $C_T$. The upwind region can be solved analytically, while the downwind region must be solved numerically using a forward marching  scheme. The model predicts a physically consistent wake evolution, with an initial flow deceleration due to pressure recovery followed by re-acceleration from turbulent entrainment, along with a gradual wake expansion. Also, the asymptotic far-wake behaviour is examined. If entrainment is driven solely by wake shear, the wake width expands as $x^{1/3}$. When background turbulence is also included, the wake expansion becomes linear with $x$ (i.e., a constant wake expansion rate). However, for small values of the incoming turbulent intensity, the wake expands linearly only at sufficiently large downstream distances, and at intermediate downwind distances, the wake experiences a variable expansion rate. Flow quantities predicted by the new actuator disk model are compared with a wide range of experimental and numerical data.   

The new theory provides a new framework to determine the $C_T$--$a$ relationship. This relation is achieved by enforcing the condition that the axial component of the net pressure force on the lateral control surface asymptotically vanishes as $x \to \infty$. A simple iterative method then yields $U$, $P$, $\sigma$, and $C_T$ as functions of $a$. The results show that for small values of $a$ or $C_T$, the dependence between $C_T$ and $a$ is largely unaffected by wake turbulent mixing, whereas for larger values of $a$ or $C_T$, it becomes highly sensitive to the level of turbulent entrainment in the near-rotor downstream wake region. The maximum power coefficient can slightly exceeds the Betz limit, and we analyse the causes for this effect. Notably, the model reduces to the Froude classical actuator disk relationship between $a$ and $C_T$ when turbulence entrainment is neglected.

As discussed in section \ref{sec:3-1}, future works could model the neglected non-linear terms in the pressure Poisson equation to have more realistic predictions of pressure $P(x)$ for highly-loaded disks. 
Future work could also investigate the applicability of the developed framework to propellers, which operate with negative induction factors. For $a < 0$, the model captures the expected velocity increase and flow contraction downwind of the propeller. However, further research is needed to validate the entrainment model and its suitability for propeller flows.

\section*{Acknowledgements}
MB acknowledges the support of the EPSRC Impact Acceleration Account at Durham University.
CM and DG acknowledge support from the National Science Foundation and the Department of Energy (via NSF grant CBET-2401013).\vspace{3 mm}

\noindent AI tools were used as a grammar checker and to proofread the text.

\section*{Declaration of Interests.} The authors report no conflict of interest. 

\begin{appen}
\section{Pressure Poisson equation for actuator disk flows}\label{sec:appendix}

Let us consider an axisymmetric, steady, incompressible flow without swirl around an actuator disk of radius \( R = D/2\), subjected to an incoming flow with velocity \( U_0 \). Note that in this Appendix, unlike the rest of the paper, our variables (e.g., $P$ and $U$) depend on both streamwise  $x$ and radial  $r$ coordinates. The flow is described using cylindrical coordinates \((x, r)\), and \( U \) and \( V \) represent the time-averaged streamwise and radial velocity components, respectively. Taking the divergence of the momentum equation (including the actuator disk force and an eddy-viscosity term for turbulence) and combining with the continuity equation, yields the mean pressure Poisson equation  which can written as 

\begin{equation}\label{eq:final_poisson equation}
\frac{1}{\rho}\nabla^2P=- \frac{1}{\rho}\Delta P_D \frac{\partial \delta(x)}{\partial x}H(R-r) + f_1(x,r)+f_2(x,r),
\end{equation}
where 
\begin{equation}
    \nabla^2=\frac{\partial^2}{\partial x^2} + \frac{1}{r}\frac{\partial}{\partial r}\left(r\frac{\partial}{\partial r}\right),
\end{equation} and $\Delta P_D=0.5\rho C_T U_0^2$ is the total pressure drop due to forcing at $x=0$. The non-homogenous terms $f_1(x,r)$ and $f_2(x,r)$ are respectively defined as 
\begin{equation}\label{eq:f(x,r)}
    f_1(x,r)=-\left(\frac{\partial U}{\partial x}\right)^2  - \frac{1}{r}\frac{\partial (rU)}{\partial r} \frac{\partial V}{\partial x},
\end{equation}
and
\begin{equation}\label{eq:g(x,r)}
   f_2(x,r)= \frac{\partial \nu_t}{\partial x} \frac{1}{r} \frac{\partial}{\partial r} \left( r \frac{\partial U}{\partial r} \right)+\frac{\partial \nu_t}{\partial r}\frac{\partial}{\partial r}\left(\frac{1}{r}\frac{\partial \left(rV\right)}{\partial r}\right).
\end{equation}

\subsection{Neglecting non-homogenous terms in pressure Poisson equation}\label{sec:neglect_f_g}
We here show that the terms $f_1$ and $f_2$ are small compared to the expected impact of the actuator disk force which must be balanced by the pressure Laplacian.
The term $f_1$ arises from the advective nonlinearity of the RANS equations (see equation \eqref{eq:f(x,r)}), while the term $f_2$ originates from spatial variations in turbulent viscosity (see equation \eqref{eq:g(x,r)}). Therefore, removing $f_1$ and $f_2$ terms is equivalent to solving the linearised RANS equations with constant turbulent viscosity. 
To estimate the order of magnitude of each term, we introduce characteristic velocity and length scales: \(\mathcal{U}_0\) represents the magnitude of the inflow velocity, while \(\mathcal{U}_d\) characterises the velocity deficit. The cross-stream and streamwise length scales are denoted by \(\mathcal{L}_r\) and \(\mathcal{L}_x\), respectively. 
The following equation expresses the order of magnitude of each term in the pressure Poisson equation:
\begin{equation}\label{eq:PPE_dimension}
\underbrace{\frac{1}{\rho}\frac{\partial^2 P}{\partial x^2}}_{\text{\large $\frac{\mathcal{U}_0\mathcal{U}_d}{\mathcal{L}_x^2}$}} + \underbrace{\frac{1}{\rho}\frac{1}{r}\frac{\partial}{\partial r}\left(r\frac{\partial P}{\partial r}\right)}_{\text{\large $\frac{\mathcal{U}_0\mathcal{U}_d}{\mathcal{L}_r^2}$}} = - \frac{1}{\rho}\Delta P_D \frac{\partial \delta(x)}{\partial x}H(R-r)  \, + 
\underbrace{f_1(x, r)}_{\text{\large $\frac{\mathcal{U}_d^2}{\mathcal{L}_x^2}$}} + 
\underbrace{f_2(x, r)}_{\text{\large $\frac{\mathcal{U}_d^2}{\mathcal{L}_x \mathcal{L}_r}$}}
\end{equation}
To estimate the pressure variation scale, we assume based on Bernoulli’s equation that $\delta P/\rho\propto  U_0 \, \delta U \sim   \mathcal{U}_0\mathcal{U}_d$  when \(\mathcal{U}_d \ll \mathcal{U}_0\). For the turbulent viscosity $\nu_t$ in $f_2(x,r)$, we assume \(\mathcal{O}(\nu_t) = \mathcal{U}_d \mathcal{L}_r\) \citep{tennekes1972first}. From Equation~\eqref{eq:PPE_dimension}, it follows that for lightly loaded disks (i.e., small values of $a$) where $\mathcal{U}_d/\mathcal{U}_0 \to 0$, the terms $f_1$ and $f_2$ become negligible compared to the terms in the Laplacian, as a leading-order approximation. In this case, the two Laplacian terms on the left-hand side must balance each other away from the disk ($\mathcal{L}_x \propto \mathcal{L}_r$) and each still much larger than $f_1$ and $f_2$ near the disk. 
Therefore, while we present the complete solution of the pressure Poisson equation in the following for the sake of completeness, we omit the effects of $f_1$ and $f_2$ for simplicity in the final result.

\subsection{Solving the Poisson equation in an infinite domain}
First, we solve 
\begin{equation}\label{eq:final_poisson equation}
\frac{1}{\rho}\nabla^2P=- \frac{1}{\rho}\Delta P_D \frac{\partial \delta(x)}{\partial x}H(R-r)  
\end{equation}  for an infinite domain $-\infty<x<\infty$ using the Greens function method. For our model we seek the solution \( P(x, r) \) evaluated at the centreline \( r = 0 \), i.e., \( P(x, 0) \). The domain of the variables is
\[
x \in (-\infty, \infty), \quad r \in [0, \infty),
\]
with the following boundary conditions:
\[
\begin{aligned}
&\lim_{x \to \pm\infty} P(x, r) = 0, \\
&\lim_{r \to \infty} P(x, r) = 0, \\
&\left. \frac{\partial P}{\partial r} \right|_{r = 0} = 0 \quad \text{(due to axisymmetry)}.
\end{aligned}
\]
In Cartesian coordinates, the general solution of $(1/\rho)\nabla^2 p= q(x,y,z)$ on $\mathbb{R}^3$, subject to zero Dirichlet conditions at infinity is:
\begin{equation}\label{infsol}
\frac{1}{\rho}p(x,y,z)=\int_{x'=-\infty}^\infty\int_{y'=-\infty}^\infty\int_{z'=-\infty}^\infty \frac{-q(x',y',z')}{4\pi\sqrt{(x-x')^2+(y-y')^2+(z-z')^2}}\, \upd z' \upd y' \upd x'.
\end{equation}
In polar coordinates, with $P(x,r,\theta)=p(x,r\cos\theta,r\sin\theta)$, $Q(x,r,\theta)=q(x,r\cos\theta,r\sin\theta)$, this amounts to
\[
\frac{1}{\rho}P(x,r,\theta)=\int_{x'=-\infty}^\infty\int_{r'=0}^\infty\int_{\theta'=0}^{2\pi} \frac{-\,r'Q(x',r',\theta')}{4\pi\sqrt{(x-x')^2+(r')^2-2rr'\cos(\theta-\theta')+r^2}}\, \upd \theta' \upd r' \upd x'.
\]
If $Q$ is axisymmetric, integrate over $\theta'$ and set $a=(x-x')^2+(r')^2+r^2,\ b=2rr',\ \phi=\theta'-\theta$ to get
\begin{align*}
\int_{\theta'=0}^{2\pi}\frac{\upd \theta'}{\sqrt{a-b\cos(\theta-\theta')}}=&\int_{\phi=-\theta}^{2\pi-\theta}\frac{\upd \phi}{\sqrt{a-b\cos\phi}}=\int_{\phi=0}^{2\pi}\frac{\upd \phi}{\sqrt{a-b\cos\phi}}=\int_{\psi=0}^{\pi}\frac{2\upd \psi}{\sqrt{a+b-2b\cos^2\psi}}\\
=&\,\frac{4K(\sigma)}{\sqrt{a+b}}\,,\quad \text{where } \sigma=\sqrt{2b/(a+b)}.
\end{align*}
Here $K(\sigma)$ is the complete elliptic integral of the first kind. For computational purposes, the identity
\begin{equation}\label{kagm}
K(\sigma)=\frac{\pi}{2 M(1,\sqrt{1-\sigma^2})}
\end{equation}
is useful, where $M(m,n)$ is the arithmetic-geometric mean (which is computed iteratively, with quadratic convergence). This yields the axisymmetric solution
\begin{equation}\label{axiinfsol}
\frac{1}{\rho}P(x,r)=\int_{x'=-\infty}^\infty\int_{r'=0}^\infty \frac{-\,r'Q(x',r')}{2\sqrt{(x-x')^2+(r+r')^2}\ \,M\!\left(1,\frac{\sqrt{(x-x')^2+(r-r')^2}}{\sqrt{(x-x')^2+(r+r')^2}}\right)}\, \upd r' \upd x'.
\end{equation}
On the axis of symmetry, \eqref{axiinfsol} reduces to
\begin{equation}\label{axiinfcl}
\frac{1}{\rho}P(x,0)=\int_{x'=-\infty}^\infty\int_{r'=0}^\infty \frac{-\,r'Q(x',r')}{2\sqrt{(x-x')^2+(r')^2}}\, \upd r' \upd x',
\end{equation}
which can also be obtained directly from \eqref{infsol} by setting $y=z=0$ and then transforming to polar coordinates.
Note that $Q(x,r)$ must decay to zero as $r\rightarrow\infty$ in order for $P(x,0)$ to be finite.

By linear superposition, each summand in the function $Q(x,r)$ can be integrated separately to give its contribution to the solution.
For instance, taking $Q(x,r)= -(1/\rho) \Delta P_D\, \delta'(x)\, H(R-r)$, where $\delta'$ denotes the formal derivative of the delta function, gives
\begin{align}\label{eq:infi_sol_f=g=0_App}
P(x,0)=&\int_{x'=-\infty}^\infty \Delta P_D\, \delta'(x')\int_{r'=0}^R \frac{r'}{2\sqrt{(x-x')^2+(r')^2}}\, \upd r' \upd x'\nonumber\\
=&\,\tfrac{1}{2}\,\Delta P_D\left(\frac{x}{\sqrt{x^2+R^2}}\,-\,\frac{x}{|x|}\right).
\end{align}
An important limitation of the infinite-domain solution is that for a first-order approximated solution where we assume $f_1 = f_2 = 0$, equation~\eqref{eq:infi_sol_f=g=0_App}  predicts a symmetric pressure discontinuity across the rotor disk, i.e., $|P_D^-| = |P_D^+| = 0.5\Delta P_D$ at $x = 0$, which is incorrect. As discussed in section \ref{sec:pressure_variation}, the pressure drop across the rotor is not symmetrical, and generally $|P_D^-|<|P_D^+|$. 
This discrepancy indicates that the non-homogenous terms in the Poisson equation must be retained for this solution to capture the asymmetric pressure jump across the disk. In this regard, the semi-infinite domain solution (i.e., solving for $x>0$ and $x<0$ separately) may be more suitable, as it allows specification of pressure right upstream or downstream of the disk from equation \eqref{eq:P_D}, offering a more realistic modelling framework even without considering the non-homogenous terms of $f_1$ and $f_2$. 

\subsection{Solving Laplace equation in two semi-infinite domains}
We here consider the domain   \( x > 0 \), i.e. starting immediately after the disk, where its forcing has no contribution and can be eliminated.
We again set $f_1 =f_2 =0$ but consider that the most important effect of $f_1+f_2$ on the pressure distribution is approximated by setting the boundary condition to the pressure's known value at the centerline immediately behind the disk, i.e., $P_{D}^-$ as specified by equation \eqref{eq:P_D}. This value is imposed as a Dirichlet boundary condition at \( x = 0 \) and $r<R$ for the Laplace equation. We thus solve: 
\begin{equation}\label{eq:PPE_smiinf}
\nabla^2P
= 0,
\end{equation}
with  boundary conditions:
\[
\begin{aligned}
P(x = 0, r) &= P_D^- \, H(R - r), \\
\lim_{x \to \infty} P(x, r) &= 0, \\
\lim_{r \to \infty} P(x, r) &= 0, \\
\left. \frac{\partial P}{\partial r} \right|_{r = 0} &= 0 \quad \text{(axisymmetry)}.
\end{aligned}
\]
where $P_D^-<0$ is a constant. Again, we are interested in the solution along the axis of symmetry, i.e., \( P(x>0, r = 0) \). In Cartesian coordinates, the general solution of $\nabla^2 p= q(x,y,z)$ on the half-space $x\geq 0$, subject to $p(0,y,z)=s(y,z)$ and zero Dirichlet conditions at infinity is:
\begin{align}
\frac{1}{\rho}p(x>0,y,z)=&\iiint \frac{q(x',y',z')}{4\pi}\left\{\frac{1}{\sqrt{(x+x')^2+(y-y')^2+(z-z')^2}}-\frac{1}{\sqrt{(x-x')^2+(y-y')^2+(z-z')^2}}\right\} \upd x' \upd y' \upd z'\nonumber\\
&+\frac{1}{\rho}\iint\frac{xs(y',z')}{2\pi}\left\{x^2+(y-y')^2+(z-z')^2\right\}^{-3/2}\upd y' \upd z'.\label{hspoisscart}
\end{align}
In terms of polar coordinates, with $S(r,\theta)=s(r\cos\theta,r\sin\theta)$ and the assumption of axisymmetry, this solution amounts to
   \begin{align}
	\frac{1}{\rho}P(x>0,r)=&\int_{x'=0}^\infty\int_{r'=0}^\infty \frac{r'Q(x',r')\,K\!\left(\frac{2\sqrt{r'r}}{\sqrt{(x+x')^2+(r+r')^2}}\right)}{\pi\sqrt{(x+x')^2+(r+r')^2}}-\frac{r'Q(x',r')\,K\!\left(\frac{2\sqrt{r'r}}{\sqrt{(x-x')^2+(r+r')^2}}\right)}{\pi\sqrt{(x-x')^2+(r+r')^2}}\, \upd r' \upd x'\nonumber\\[4pt]
	&+\frac{1}{\rho}\int_{r'=0}^\infty\frac{2x\,r'S(r')\,E\!\left(\frac{2\sqrt{r'r}}{\sqrt{x^2+(r+r')^2}}\right)}{\pi\sqrt{x^2+(r+r')^2}\,\{x^2+(r-r')^2\}}\, \upd r',
\end{align}
where $E$ is the complete elliptic integral of the second kind. Here, if we assume the sum of non-homogenous terms in equation \eqref{eq:PPE_smiinf} is zero (i.e. $Q(x,r)=f_1(x,r)+f_2(x,r)=0$), the solution simplifies to 
  \begin{equation}\label{eq:P(x,r)}
P(x>0,r)=\int_{r'=0}^\infty\frac{2x\,r'S(r')\,E\!\left(\frac{2\sqrt{r'r}}{\sqrt{x^2+(r+r')^2}}\right)}{\pi\sqrt{x^2+(r+r')^2}\,\{x^2+(r-r')^2\}}\, \upd r'.
\end{equation}
Figure \ref{fig:radial_pressure} shows the variation of $P(x,r)$, computed from equation \eqref{eq:P(x,r)}, at several streamwise locations downstream of the disk. Immediately behind the disk, where the pressure is relatively high, the radial pressure profiles exhibit a top-hat shape. Further downstream, the pressure drops sharply, and the profiles become increasingly smooth. For simplicity, here we use the pressure along the axis of symmetry as a representative measure of how pressure evolves with streamwise position. This provides the necessary closure relation for the actuator disk theory developed in the main part of the paper. On the axis of symmetry, equation \eqref{eq:P(x,r)} simplifies to
\begin{equation}\label{eq:PPE_solution_semiinf_f=g=0_x>0}
P(x>0,0)=\int_{r'=0}^R\frac{P_D^{-}x\,r'}{\{x^2+(r')^2\}^{3/2}} \upd r'=P_D^{-}\left(1-\frac{x}{\sqrt{x^2+R^2}}\right). 
\end{equation}
\begin{figure}
    \centering
    \includegraphics[width=0.5\linewidth]{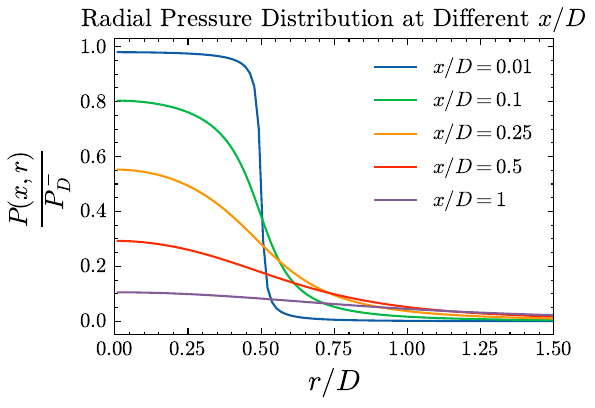}
    \caption{Profiles of $P(x,r)$ normalised by $P_D^-$ based on equation \eqref{eq:P(x,r)}}
    \label{fig:radial_pressure}
\end{figure}

As an alternate method of solving the Laplace equation, the same result can be also obtained using the Hankel transform method. For the $x>0$ domain, this leads to $P(x,r) = P_D^{+} \, R \int_0^\infty J_1(\rho R) \,J_0(\rho r) \, e^{-\rho x} \, \upd\rho $, where $J_0$ and $J_1$ are zero and first order Bessel functions. Evaluating at $r=0$ and performing the integral of the exponentials-weighted $J_1$ function yields the same result as in equation \eqref{eq:PPE_solution_semiinf_f=g=0_x>0}.

Equation~\eqref{eq:PPE_solution_semiinf_f=g=0_x>0} can also be applied to the upwind region by substituting \(-x\) for \(x\), and using \(P_D^+\) as the corresponding boundary condition. Thus, the solution for \(P(x, 0)\) over the entire domain \( -\infty < x < \infty \), $x\neq 0$, assuming $f_1=f_2=0$, is given by:
\begin{equation}\label{eq:PPE_solution_semiinf_f=g=0_app}
   P(x,0)=P_D^{+}\left(1+\frac{x}{\sqrt{x^2+R^2}}\right)\mathrm{H}(-x)+P_D^{-}\left(1-\frac{x}{\sqrt{x^2+R^2}}\right)\mathrm{H}(x),
\end{equation}

We have replaced the actual effects of $f_1+f_2$ with specification of the distinct values of $P_D^+$ and $P_D^-$ that are obtained instead from an ``integral'' condition (integrated Bernoulli equation) of the problem. Finally, it is worth noting that the simplified solution presented here does not capture the complexities of the pressure distribution around the disk or outside the control volume, and should be regarded only as an approximation of pressure variation along the disk axis. Finally, it is worth mentioning that for small perturbations, assuming the pressure satisfies Laplace's equation ($\nabla^2 p = 0$) is 
equivalent to using potential flow theory with Laplace's equation for the velocity potential ($\nabla^2 \phi = 0$), where objects can be modelled as sources or sinks 
(similar to the approach used in \cite{steiros_drag_2018}).

\end{appen}\clearpage

\bibliographystyle{jfm}
\bibliography{ref.bib,references_Zotero}

\end{document}